\documentclass[prc,nofootinbib,showpacs,twocolumn]{revtex4}

\usepackage{graphicx}
\usepackage[usenames,dvipsnames]{color}
\usepackage{amsmath,amssymb,bbold}
\usepackage{hyperref}

\begin{document}
\title{Effects of magnetic field on the plasma evolution in 
relativistic heavy-ion collisions}
%

%
\author{Arpan Das$^{1,2}$ \footnote {email: arpan@iopb.res.in}}
\author{Shreyansh S. Dave$^{1,2}$ \footnote {email: shreyansh@iopb.res.in}}
\author{P.S. Saumia$^3$ \footnote {email: saumia@theor.jinr.ru}}
\author{Ajit M. Srivastava$^{1,2}$ \footnote{email: ajit@iopb.res.in}}
\affiliation{$^1$ Institute of Physics, Bhubaneswar 751005, India\\
$^2$ Homi Bhabha National Institute, Training School Complex,
Anushakti Nagar, Mumbai 400085, India\\
$^3$ Bogoliubov Laboratory of Theoretical Physics, JINR,
141980, Dubna, Russia }

\begin{abstract}

Very strong magnetic fields can arise in non-central heavy-ion
collisions at ultrarelativistic energies, which may not decay quickly in a
conducting plasma. We carry out relativistic magnetohydrodynamics (RMHD)
simulations to study the effects of this magnetic field on the evolution
of the plasma and on resulting flow fluctuations in the ideal RMHD limit. 
Our results show that magnetic field leads to enhancement in elliptic flow
for small impact parameters while it suppresses it for large impact 
parameters (which may provide a signal for initial stage magnetic field). 
Interestingly, 
we find that magnetic field in localized regions can temporarily increase 
in time as evolving plasma energy density fluctuations lead to reorganization 
of magnetic flux. This can have important effects on chiral magnetic 
effect. Magnetic field has non-trivial effects on the power spectrum of
flow fluctuations. For very strong magnetic field case one sees a pattern
of even-odd difference in the power spectrum of flow coefficients arising
from reflection symmetry about the magnetic field direction if initial
state fluctuations are not dominant.  We discuss the situation of 
nontrivial magnetic field configurations arising from collision of 
deformed nuclei and show that it can lead to anomalous elliptic flow. 
Special (crossed body-body) configurations of deformed nuclei
collision can lead to presence of quadrupolar magnetic field which can
have very important effects on the rapidity dependence of transverse expansion
(similar to {\it beam focusing} from quadrupole fields in accelerators).
\end{abstract}
\maketitle

\section{INTRODUCTION}

Extensive efforts have focused on the discovery of the quark-gluon 
plasma (QGP) phase of QCD in relativistic heavy-ion collision 
experiments (RHICE). There is mounting evidence that QGP phase is
created in these experiments. It is no more possible to explain the
wealth of experimental data at RHIC and LHC without assuming a transient
phase of QGP. While it is certainly desirable to have {\it smoking gun}
signal for QGP, it is also an appropriate stage for the exploration of
the rich spectrum of physics unfolded by the (most likely) presence
of this transient stage of QGP in relativistic heavy-ion collisions.
Search for exciting possibilities like the critical point in the QCD
phase diagram, possible exotic high baryon density phases (in upcoming
facilities FAIR and NICA) are some of these directions.

  An entire new line of explorations has been initiated in recent
years by the very exciting possibility that in relativistic heavy-ion
collision experiments extremely high magnetic fields are expected to
arise, especially in non-central collisions. During earliest stages,
magnetic field in the plasma can be of order 10$^{15}$ Tesla (few $m_\pi^2$), 
which is several orders of magnitude larger than the magnetic field even in 
magnetars. Such a strong magnetic field in QGP will lead to important effects.
Much of the discussion in literature has focused on the exciting possibility 
of observing CP violation effects \cite{cp}. Relevant
effects are generally known as {\it chiral magnetic effect} and more
recently discussed {\it chiral vortical effect}. Along with such effects,
there are many other important consequences of the magnetic field for
QGP evolution. Some of us had earlier utilized the fact that
an important effect of the presence of magnetic fields in the 
plasma will be to lead to variations in velocities of different 
types of waves in the plasma \cite{bv2}. In particular the group velocity 
varies with the angle between the wave vector and the direction of
the magnetic field. Its obvious effect will be to qualitatively modify
the development of anisotropic flow. In ref. \cite{bv2}, it was argued
that the flow coefficients can be significantly affected by these effects, 
in particular, the presence of magnetic field can lead to enhancement in the 
elliptic flow coefficient $v_2$ by almost 30\%. As pointed out in
ref. \cite{bv2}, it  raises the interesting possibility of whether a
larger value of $\eta/s$ can be accommodated by RHIC data when these
effects are incorporated using full magnetohydrodynamical simulations.
This possibility can be viewed as either leading to the QGP $\eta/s$
being higher than the AdS/CFT bound, or in the context of results in
ref.\cite{etabys} which suggested crossing the AdS/CFT bound, to restore
the bound when proper effects of magnetic field are incorporated. 
The issue of magnetic field dependence of elliptic flow was discussed
by Tuchin \cite{tuchinv2} (including viscous effects as well) with 
results in agreement with \cite{bv2}.

 The arguments in ref.\cite{bv2} utilized directional dependence of
sound velocity in the presence of magnetic field and modeled its effect
on development of elliptic flow. Those results were not based
on any magnetohydrodynamical simulation. In this paper, we have
carried out detailed relativistic magnetohydrodynamics simulations.
We indeed confirm the results in \cite{bv2,tuchinv2} that elliptic flow
can increase in the presence of magnetic field. However, our results
show that the dependence of $v_2$ on magnetic field is much more complex
than assumed in these earlier works, with several factors at play. In certain
situations (e.g. for small impact parameters) the magnetic field enhances
the elliptic flow, while in a different situation (large impact parameter),
magnetic field suppresses the elliptic flow. These underlying factors
are important to understand (especially in view of other recent 
relativistic magnetohydrodynamics simulations \cite{mhdv2} where 
it was found that magnetic field has no effect on elliptic flow, in
contrast to the results in \cite{bv2,tuchinv2}). Along with the effect on 
elliptic flow, we will demonstrate several other important effects
of magnetic field showing how  flow evolution is qualitatively affected.
These effects are important as they show that an understanding of
flow pattern is not complete without including effects of magnetic field
in the early stages of plasma evolution. 

 Another important reason to focus on different qualitative effects
of magnetic field on flow evolution is that it can provide signal for the 
presence of strong magnetic field during early stages of the plasma 
evolution. It must be emphasized that although for the earliest stage of
collision, magnetic field can be calculated with reasonably accurate
approximations, its evolution even in immediately successive stages
is poorly understood. All the important effects of the magnetic field,
such as chiral magnetic effect as well as
various effects on flow pattern as we have discussed here (and in
\cite{bv2,tuchinv2}) require that reasonably strong magnetic field
survives for at least several fm proper time. Earlier it was thought
that magnetic field rapidly decays after the collision. It is strong
for a very short time, essentially the passing time of the Lorentz 
contracted nuclei ($\sim$ 0.2 fm for RHIC energies). Subsequently it 
rapidly decays  \cite{cp2,bcal1}.
In such a situation the effect of magnetic field on flow evolution
as well effects such as chiral magnetic effect 
will be strongly suppressed as time scale for the development of flow 
and for charge separation (for latter effects) is several fm. Similar
situation is expected at higher energies, e.g. at LHC.

 It was later pointed out by Tuchin \cite{tuchin} that
magnetic field does not decay very rapidly due to induced currents arising
from rapidly decreasing external magnetic field. In fact, the
magnetic field satisfies a diffusion equation with the diffusion
constant equal to $1/(\sigma \mu)$ where $\mu$ is the magnetic permeability
and $\sigma$ is the electrical conductivity \cite{dfusn,jcksn}. 
With $\mu \sim 1$ and $\sigma \simeq 0.04 T$ (= 0.04 fm$^{-1}$ for 
T $\simeq$ 200 MeV) from refs.\cite{sigma}, one finds that the
time scale $\tau$ over which the magnetic field remains reasonably strong
\cite{tuchin} over length scale $L$ is,
$\tau \simeq L^2 \sigma/4$. For $L = 6 - 10$ fm, we get $\tau$ less than
1 fm. Indeed, one sees that magnetic field decreases fast initially, 
though at later times matter effects become more important slowing down
decrease of magnetic field significantly \cite{bref}.  
For higher temperatures $\sigma$ 
will be larger increasing the value of $\tau$. $\sigma$ is also expected 
to increase due to the  effects of magnetic
field in the plasma \cite{sigmaB}, further increasing the value of $\tau$.

 Even if one takes this optimistic picture that due to non-zero
conductivity of QGP, magnetic field doe not decay extremely rapidly,
and may survive for significant
time scales, the self-consistency of this picture can be questioned
due to uncertainties of the initial non-equilibrium stages of
the parton system. Initially there is no plasma, so no conductivity.
If the parton system is assumed to have the QGP conductivity during
its formation stages, question arises as to how magnetic field
can penetrate the conducting plasma. For simplicity consider the 
plasma (say during early stage for less than 1 fm lab time) to be a 
thick disk of nuclear diameter and thickness of 1-2 fm. If the plasma
was static then one could just consider the penetration depth 
$\delta \sim (\mu \sigma \omega)^{-1/2}$ where $\omega$ is the angular 
frequency of electromagnetic wave.
Initial magnetic field, being a narrow pulse of time duration
$t \simeq 0.2$ fm (typically the width of Lorentz contracted Nuclei,
for RHIC energies), can be taken to have $\omega \simeq 30$ fm$^{-1}$.
This gives the penetration depth of order 1 fm (note it was
mistakenly written as 3 fm in ref.\cite{bv2}). In such a situation, 
though magnetic field cannot penetrate from the perimeter of the
disc (nuclear radius being about 6-7 fm), it may be able to penetrate
significantly in the interior from the longitudinal direction (from both 
sides), with disk thickness being only about 1 fm. In such a situation, 
the picture of magnetic field diffusing through the entire region of the plasma
with typical length scale of several fm, and lasting with a value close to
the high initial peak values for time scales of several fm, may be
self consistent.

  However, the plasma is not static in the longitudinal direction.
Far from it, the plasma is relativistically expanding in the longitudinal
direction. The above argument of penetration depth cannot be applied
to a conducting plasma which is relativistically expanding. The conclusion
being that if the plasma is taken to be conducting from the very beginning,
we do not know how much fraction of the original magnetic field penetrates 
the plasma. Only that fraction can then be assumed to follow the
diffusion equation as in ref.\cite{tuchin} and survive for few fm
time scale. A proper treatment of the problem will require treatment of
the early parton system as a non-equilibrium system, whose response
to ambient magnetic field then needs to be estimated. As the plasma
equilibrates, it will develop conductivity as appropriate for the
QGP phase, and one needs to determine how much magnetic field is
trapped in the conducting plasma. Its subsequent evolution then can
be followed as in ref.\cite{tuchin}. 

 Having stated all these issues, we will take a simple path. We will
assume, for simplicity, that strong magnetic field exists inside the 
plasma. The strength
of magnetic field will be estimated close to its peak value, and will be 
assumed to survive in the plasma for the duration of evolution we carry out
evolving according to the equations of relativistic magnetohydrodynamics
(RMHD).  We assume ideal RMHD with infinite conductivity, so
magnetic field lines are frozen in the plasma. Due to our computer 
limitations (for our 3+1 dimensional simulations), we are only able to 
consider small lattice, hence evolve for short times
up to about 3 fm (to avoid boundary effects). This being a rather short 
time, our assumption of ideal MHD may not be very inappropriate.
Our focus in the paper is mainly on the qualitative aspects of
results, and not on actual numbers. We are not claiming to give
numbers which can be compared to the experimental data. Rather we 
demonstrate qualitative patterns of flow evolution, which one can look for
in the experiments.  Primary 
emphasis being on these being signals of the presence of  strong
magnetic field during early stages of plasma evolution.

 The paper is organized in the following manner. In Sec.II, we 
briefly review the formalism we have adopted for the relativistic 
MHD simulation from ref.\cite{mhd}. Sec.III presents details of
our numerical simulation. In Sec.IV we first discuss the issue of 
effect of magnetic field on elliptic flow
(in view of conflicting conclusions in \cite{bv2,tuchinv2,mhdv2}).
Sec.V presents results of the simulations which show that magnetic field
can lead to qualitatively new effects. Sec.VI presents conclusions
and discussions.

\section{THE FORMALISM}

 We here provide a brief summary of the formalism we have followed for
our relativistic magnetohydrodynamical (RMHD) simulations. For this we
have followed ref.\cite{mhd} and for the benefit of the reader we provide 
essential steps from that ref. in the following. We will be assuming zero
baryon chemical potential situation so there will not be any baryon 
number conservation equation. Equations for ideal RMHD
for the evolution of fluid and magnetic field are as follows.

Conservation of total energy momentum tensor (for QGP as well as
the magnetic field) is given by

\begin{equation}
\partial_\alpha [(\rho + p_g + |b|^2) u^\alpha u^\beta -
b^\alpha b^\beta + (p_g + {|b|^2 \over 2}) \eta^{\alpha \beta}] = 0 ,
\end{equation}

where we have used perfect fluid form for the QGP energy-momentum
 tensor,

\begin{equation}
T^{\mu \nu} = (\rho + p_g)u^\mu u^\nu + p_g \eta^{\mu \nu} .
\end{equation}

Maxwell's equations are

\begin{equation}
\partial_\alpha(u^\alpha b^\beta - b^\alpha u^\beta) = 0 .
\end{equation}

Here $\rho$ and $p_g$ are the energy density and pressure of
QGP which we assume to be related by an ideal gas equation of state,
$p_g = {\rho \over 3}$. $u^\alpha$ is the four-velocity with
$u^\alpha u_\alpha = -1$. Four-vector $b^\alpha$ is related to
the magnetic field ${\vec B}$ as,

\begin{equation}
b^\alpha = \gamma [{\vec v}.{\vec B} , {{\vec B} \over \gamma^2}
+ {\vec v} ({\vec v}.{\vec B})] .
\end{equation}

$\gamma$ is the Lorentz factor for velocity ${\vec v}$ and we have
following normalizations

\begin{equation}
u^\alpha b_\alpha = 0, ~{\rm and}~ |b|^2 \equiv b^\alpha b_\alpha
= {|{\vec B}|^2 \over \gamma^2} + ({\vec v}.{\vec B})^2 .
\end{equation}

For numerical simulation, the above equations are cast in the following
form

\begin{equation}
{\partial U \over \partial t}  + \sum_k {\partial F^k \over
\partial x^k} = 0 .
\end{equation}

This is the evolution equation for the vector $U$ where

\begin{equation}
U = (m_x, m_y, m_z, B_x, B_y, B_z, E) ,
\end{equation}

where 
\begin{equation}
m_k = [\rho h \gamma^2 + |{\vec B}|^2]v_k - ({\vec v}.{\vec B})
B_k ,
\end{equation}
and
\begin{equation}
E = \rho h \gamma^2  - p_g + {|{\vec B}|^2 \over 2}
+ {v^2 |{\vec B}|^2 - ({\vec v}.{\vec B})^2 \over 2} .
\end{equation}

$h$ is the specific enthalpy ($4p_g/3$ with the ideal gas equation of
state we are using for QGP) and $F^k$ are the fluxes in Eqn.(6) 
along directions $x^k \equiv (x,y,z)$  given as follows.
\begin{equation}
F^x = \begin{pmatrix}m_xv_x - B_x {b_x \over \gamma} + p\\
m_yv_x - B_x {b_y \over \gamma}\\
m_zv_x - B_x {b_z \over \gamma}\\ 0 \\
B_yv_x - B_xv_y\\
B_zv_x - B_xv_z\\ m_x\end{pmatrix} .
\end{equation}

Similar expressions hold for $F^y$ and $F^z$ by appropriate
replacement of indices. Here $p = p_g + {|b|^2 \over 2}$ is the total
pressure. Evolution is carried out using Eqn.(6) for the vector $U$ 
from which the independent variables $(p_g, {\vec v}, {\vec B})$ have to
be extracted. For this we define $W = \rho h \gamma^2$ and
$S = {\vec m}.{\vec B}$. With this we can write,

\begin{equation}
E = W - p_g + (1 - {1 \over 2\gamma^2}) |{\vec B}|^2 - {S^2 \over 2W^2} ,
\end{equation}

\begin{equation}
|m|^2 = (W + |{\vec B}|^2)^2 (1 - {1 \over \gamma^2}) - {S^2
\over W^2} (2W + |{\vec B}|^2) .
\end{equation}

This equation is used to express $\gamma$ as a function of $W$
and known variables ${\vec m}$, ${\vec B}$, and hence $S$
(from the knowledge of vector $U$).

\begin{equation}
\gamma = \left( 1 - {S^2(2W + |{\vec B}|^2) + |m|^2 W^2 \over
(W + |{\vec B}|^2)^2 W^2} \right)^{-1/2} .
\end{equation}

With the ideal gas equation of state we have $p_g = {W \over 4 \gamma^2}$.
Eqn.(9) then can be entirely written in terms of unknown quantity $W$
and other known quantities ${\vec B}$, $S$ and $E$ as follows

\begin{equation}
f(W) \equiv  W - p_g + (1 - {1 \over 2\gamma^2}) |{\vec B}|^2 - 
{S^2 \over 2W^2} - E = 0 .
\end{equation}

We solve this equation using Newton-Raphson method to get $W$
using expressions for various derivatives as in ref.\cite{mhd}.
(Except for one derivative, we obtain $dp_g/dW$ using the equation
$p_g = {W \over 4 \gamma^2}$ for our choice of equation of state.
This expression differs from the expression in ref.\cite{mhd}.). 
From the value of W thus obtained, $\gamma$ can be calculated using 
Eqn.(13). With this, we get value of $p_g$. Equation for $m_k$
(Eqn.(8)) can then be used to obtain velocity components $v_k$.
This completes the procedure of recovery of independent variables
from time evolved vector $U$.  For further details, we refer to
ref.\cite{mhd}.

\section{DETAILS OF NUMERICAL SIMULATION}

We have developed a 3+1 dimensional code and use lattice of size
$200 \times 200 \times 200$ (and in some cases, for example for power 
spectrum for very strong magnetic field case, to get averages over
several events we use smaller lattice $150 \times 150 \times 150$).
For evolution we use leapfrog algorithm of 2nd order accuracy. Due
to small size of the lattice (due to computer limitations) we are
able to evolve only for times up to 3 fm to avoid boundary effects.
In some cases we evolve for shorter times as we will mention for
the respective cases.

Glauber like initial conditions are used for the initial energy density
profile  where a nucleus-nucleus collision is viewed as a sequence of 
independent binary nucleon-nucleon collisions \cite{gbmc}. 
The enhancement of $v_2$ is studied using smooth Glauber initial
conditions in the X-Y plane. The parameters are tuned to an initial 
central temperature of 160 - 180 MeV assuming energy density
of ideal gas of quarks and gluons for the two flavor case
with zero chemical potential. A smooth
Woods-Saxon profile is used along z-axis with extent equal to the extent of
the colliding region along Y-axis. While studying the other  effects like
flux re-organisation, any possibility of vorticity generation, 
and the power spectrum, we use Glauber Monte Carlo initial conditions 
with parameters tuned to obtain the desired temperature range of 160 - 
180 MeV. We distribute the energy density from the collision of 
participants along z-axis following a Gaussian distribution. 
As we mentioned above, 
ideal gas equation of state is used for QGP. We also add a constant 
background energy density of about 1\% of the maximum initial energy
density of the plasma, it gave better stability for the simulation,
especially in the presence of fluctuations. (This energy density addition 
is not needed due to any instability of the program. Our algorithm of 
extracting the primitive variable does not work effectively when 
magnetic field energy density is much larger than the plasma density,
as was noted in ref.6 also. Hence a non-zero energy density is used in 
the outer region. Such a small energy density should not affect any 
results strongly. Indeed, it is hard to argue that surroundings of the 
QGP region do not have some small energy density.)
For some of the results, we have
neglected fluctuations and have used Glauber optical initial conditions
along the x-y plane.
This is done so that one can isolate the effects of magnetic field
on the specific features of plasma evolution. Fluctuations lead to
magnetic flux rearrangement which makes evolution highly complex,
as we will demonstrate in the section on results. So, in the presence of
fluctuations it becomes hard to associate specific patterns of flow
with the magnetic field. Certainly, for experimental comparison
one will need to combine all the effects together, and make special
efforts to identify specific regimes of collision energy, nuclear
size, centrality etc. to enhance the effects due to magnetic field.
When we present results we specify where fluctuations are included
and where not.

 For the initial configuration of magnetic field we have used several
methods. For most of the results we use magnetic field produced by
two oppositely moving uniformly charged spheres (representing colliding 
nuclei), in vacuum, with appropriate Lorentz gamma factor \cite{jcksn,bref}.
This neglects modifications due to participants, but that is not expected  
to be very significant.  This works fine with the range of magnetic 
fields expected in relativistic heavy-ion collisions, though we restrict 
our simulation to lower energy collisions about $\sqrt{s}$ = 20 GeV. 
For most of the simulations below, the magnetic field profile is obtained
in this manner. We typically give two sets of results, labeled by 
$B_{time}$ which is the time at which the magnetic field profile
is calculated after the collision. We use $B_{time}$ = 0.4 fm and 0.6 fm.
Smaller value gives larger value of the magnetic field, but may not be 
very realistic in view of finite conductivity of the plasma. 
If we use very large Lorentz gamma factor, then the
magnetic field is sharply peaked at receding (Lorentz contracted) nuclei,
and our 3+1 dimensional simulation is not able to run for reasonable
times, especially in the presence of fluctuations. For some cases
just to show some interesting effects (e.g. systematic
difference in the power of even-odd flow coefficients) we needed to
use very high magnetic fields (of order 15 $m_\pi^2$). Such large
magnetic field are completely unrealistic here, and we use this
value only to show how completely new effects may arise for very
large magnetic field.
The simulation with realistic magnetic field profile develops
instabilities, primarily because in this case magnetic field energy
density is much larger than the plasma density everywhere. Such 
difficulties have been noticed in other simulations as well \cite{mhdv2}
where it is mentioned that the numerical code was not able to handle 
configurations where the magnetic pressure is much larger than the
thermal pressure, which typically is the case in regions outside the 
plasma region. To avoid these difficulties, for the large magnetic field 
case, we use a simpler profile for magnetic field
where the profile in the (x-z) plane is chosen to be proportional to
the energy density profile in the (x-z) plane at y = 0 obtained from
the Glauber Monte Carlo like procedure as described above.  
z axis is the collision axis and the impact 
parameter is along the x axis, with resulting magnetic field pointing 
along the y axis.  The peak value of the magnetic field is
chosen by hand. The magnetic field is then taken to be constant along
the y axis, as consistent with Gauss' law. Clearly this magnetic field
profile is not realistic along the y axis, but is only used to illustrate 
special effects of magnetic field on plasma evolution. The possibility
remains that large magnetic fields may not be very unrealistic for 
example for deformed nucleus case. 

 We mention here that for low energy collisions with $\sqrt{s}$ = 20 GeV 
 it is not appropriate to work with the simple zero chemical
 potential ideal relativistic gas  equation of state which we have used. 
 Also, at such low energies, chemical potential is sizeable and
 one cannot ignore baryon current. We use these approximations (simple
equation of state and zero chemical potential) for simplicity, just
as we have used ideal MHD equations for the evolution of the plasma.
Our aim in this work is not to give definite numbers which can
be compared with the experiments. Rather we  look for basic physics 
for new effects. These qualitative patterns will not be expected to
depend on the presence (or absence) of baryon current, or on the exact
nature of the equation of state, though the numerical values will
certainly depend on the factors. We thus expect that the qualitative 
patterns we find and the basic physics of new phenomena we discuss,
will apply from low energy collisions (e.g. at FAIR and NICA) to 
high energy collisions at LHC.

 As the simulation is carried out using (x,y,z) coordinates, with
complete 3-dimensional profile for the initial energy density and
magnetic field, we incorporate longitudinal expansion by assuming
a maximum value of the velocity (of 0.7) at maximum z value for the Lorentz
contracted energy density profile. (Note that this maximum velocity 
represents the velocity of the equilibrated plasma, and not that of
the receding nuclei.)  For intermediate distances, velocity
is assumed to vary linearly as appropriate for Bjorken scaling. We
use the lab time coordinate for time evolution. For the initial 
energy density profile, we first assume energy density profile
as appropriate for an initial constant proper time hypersurface, evolving
locally by longitudinal Bjorken scaling law, and then transform
it to the constant lab time. This neglects nonlinear effects of 
inhomogeneities on evolution for a very short proper time period
(the initial time for the beginning of plasma evolution), but 
should not be important for later time evolution. All our results are
for the central rapidity region with unit rapidity window (suitably 
translated to $\Delta z$). This further makes our results reasonably 
reliable as the difference between the proper time and lab time
are significant only for larger rapidities. Due to limitation of lattice 
size we have only considered small nucleus, copper in our case. Even for that,
we have taken the radius to vary from about 3 fm to 4.5 fm (depending 
on consideration of fluctuations etc.). We again emphasize that the
spirit of our work here is to demonstrate various important qualitative
patterns in the flow in the presence of magnetic fields, rather than
precise numbers.

\section{EFFECT OF MAGNETIC FIELD ON ELLIPTIC FLOW}

 Before we present results of our simulations for different aspects of
flow evolution, including the elliptic flow, we discuss previous results
in the literature regarding effects of magnetic field on elliptic flow.
In an earlier paper \cite{bv2}, some of us had argued that magnetic field
can lead to enhancement of elliptic flow by up to about 30\%. We first
briefly recall physical arguments for such an enhancement as discussed
in ref.\cite{bv2}. Basic argument in \cite{bv2} relied on the the effects of
an external magnetic field on sound waves in QGP produced
in RHICE. For relativistic magnetohydrodynamics, the waves which are relevant 
for the case of discussion of flow are the {\it magnetosonic waves}
as they involve density perturbations. Phase velocities for these
waves are given by \cite{mhdwave}

\begin{equation}
{\bf v}_{ph} = v_{ph} {\bf n} = {\bf n} ({1 \over 2}[(\rho_0 h_0 /\omega_0)
c_s^2 + v_A^2])^{1/2} (1 + \delta \cos^2\theta \pm a)^{1/2} .
\end{equation}

Here $+$ and $-$ signs correspond to the fast and slow magnetosonic
waves respectively, $v_A = B_0/\sqrt{\omega_0}$ is the Alfv\'en speed,
and $\delta$ and $a$ are defined below. Mean local values of various quantities
are denoted by subscript $o$ and $\theta$ is the angle between the
magnetic field and ${\bf n}$.

\begin{equation}
a^2 = (1 + \delta \cos^2\theta)^2 - \sigma \cos^2\theta , \\
\end{equation}

\begin{equation}
\delta = {c_s^2 v_A^2 \over [(\rho_0 h_0/\omega_0)c_s^2 + v_A^2]} , 
~~ \sigma = {4c_s^2 v_A^2 \over [(\rho_0 h_0/\omega_0)c_s^2 + v_A^2]^2} .\\
\end{equation}

(Note, $\sigma$ is defined above, and should not be confused with
the conductivity as used above.)
For propagation of density perturbations, as relevant for the
evolution of flow anisotropies, the relevant wave velocity is the group
velocity for the magnetosonic waves,

\begin{equation}
{\bf v}_{gr} = v_{ph}\left[{\bf n} \pm {\bf t} {[\sigma \mp 2\delta (a
\pm (1 + \delta \cos^2\theta))]\sin\theta \cos\theta \over 2(1 +
\delta \cos^2\theta \pm a)a}\right] .
\end{equation}

Here ${\bf t} = [({\bf B_0}/B_0) \times {\bf n}] \times {\bf n}$, and
again the upper and lower signs ($\pm$ or $\mp$) correspond to the
fast and the slow magnetosonic waves respectively. For a given magnetic
field ${\bf B_0}$, the direction of ${\bf n}$ can be varied to generate
group velocities of these waves in different directions. Fig.1 shows
a typical situation of various vectors in Eq.(18) expected in RHICE.
It is important to note that the direction of ${\bf v}_{gr}$ depends
on the relative factors multiplying ${\bf n}$ and ${\bf t}$ in Eq.(18).
This in turn depends on properties of the plasma like energy density.
Thus due to the presence of spatial gradients in RHICE, especially
due to initial state fluctuations, even along a
fixed azimuthal direction, we expect the direction of
${\bf v}_{gr}$ to keep varying with the radial distance. This can lead
to the development of very complex flow patterns. This raises a very
interesting possibly of generation of vorticity in the plasma entirely from
the effect of magnetic field. We are not able to fully explore this
possibility as yet due to our limitation of relatively small time
evolution of the plasma. (Vorticity will  be expected to arise at later
times when the flow pattern gets significantly twisted due to magnetic
field effects.) Any such vorticity will have important implications,
especially in view of chiral vortical effect.

\begin{figure}[!htp]
\begin{center}
\includegraphics[width=0.25\textwidth]{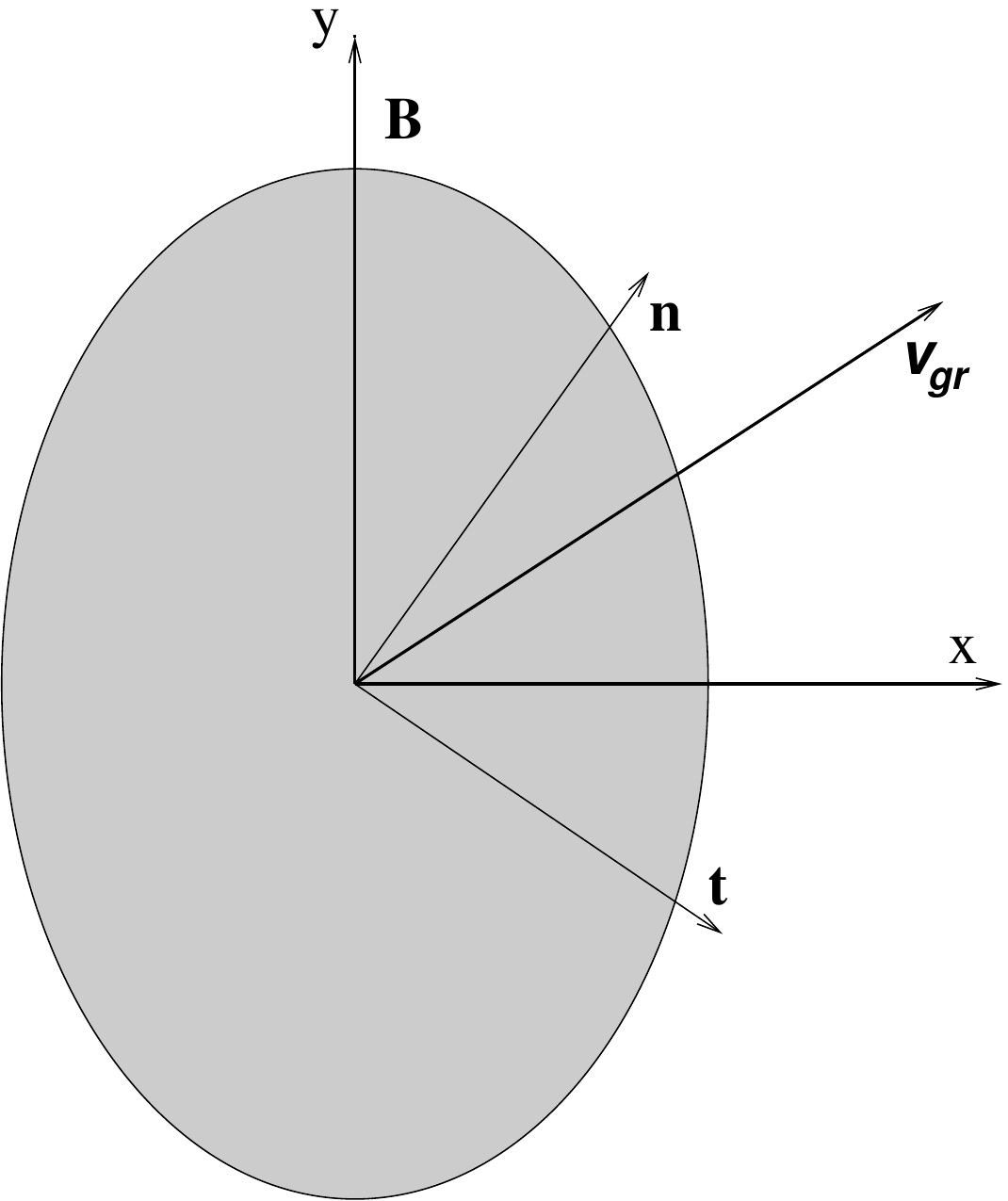}
\caption{A typical situation expected in relativistic heavy-ion collisions
with the magnetic field pointing in the y direction. The direction of the 
group velocity ${\bf v}_{gr}$ is obtained from ${\bf n}$ and ${\bf t}$ 
via Eq.(18). (figure taken from ref.\cite{bv2}).}
\label{fig1}
\end{center}
\end{figure}

The effect of magnetic field on propagation of sound waves 
here comes from an effective magnetic pressure arising from the freezing 
of magnetic field lines in the plasma in the magnetohydrodynamics limit.
The distortions of magnetic field lines in the presence of density 
perturbations cost energy leading to an extra contribution to pressure from
the presence of magnetic field. This is what is responsible for
increasing the effective sound speed as given above. The estimate
of the effect of magnetic field on elliptic flow in ref.\cite{bv2}
was based on the fact that the flow coefficients are proportional to 
the sound velocity \cite{oltr}, which now becomes dependent on
the directions of the magnetic field and that of the phase velocity.
This directly affects the flow pattern and hence elliptic flow.

  We mention that these arguments are rather crude. Elliptic flow
is a complex phenomenon and cannot be directly related to the anisotropy
of the stiffness of equation of state and resulting sound velocity.
Our intention here is to point out the underlying physics of the
phenomena and why one may expect an increased elliptic flow from
the presence of magnetic field.
A more detailed analysis of the effects of magnetic field on elliptic
flow was carried out by Tuchin in \cite{tuchinv2} with results in
agreement with the estimates of \cite{bv2}. Later, in the section of
results we will present results of our numerical simulation where
again magnetic field is found to enhance elliptic flow.
However, quite different results are reported in a recent numerical 
RMHD simulations where magnetic field was found to have no effect on 
elliptic flow \cite{mhdv2}. It is important to understand possible reasons
for the discrepancies between these different works. For this purpose
we have carried out simulations to study elliptic flow evolution with
different values of impact parameters which lead to different
types of magnetic field profiles. Our conclusion is that in the end
the effects of magnetic field on flow pattern has many complex features.
The picture used in \cite{bv2} was indeed too simplistic where the
magnetic field dependent sound speed was directly assumed to affect
the elliptic flow. In fact quite opposite arguments could be given
using Lenz's law from which one expects that induced magnetic fields
will always oppose the change which causes magnetic flux changes.
Basically this should imply that expansion along x axis should be
suppressed as this leads to decrease in magnetic flux, while expansion
along y axis should not be affected, thereby decreasing elliptic flow.
The actual situation is much more complex. For example, Lenz's law
argument does not distinguish between uniform expansion along x axis
and the distortion of a localized plasma by transverse expansion. The
latter leads to distortion of field lines, and not just decrease in 
magnetic flux, which has implications for extra pressure, and hence on 
sound waves. Some of the complexities have been discussed recently in 
refs.\cite{bv2ref1,bv2ref2}, though exact time dependence used for
magnetic field in ref.\cite{bv2ref2} seems difficult to justify,
(also for the Gaussian profile of the magnetic field in the x-y plane
in ref.\cite{bv2ref1}, one needs to ensure that Gauss' law is satisfied.)
 
  As we will see later, net effect of magnetic field on elliptic flow
depends very sensitively on the profile of magnetic field in relation
to the profile of plasma energy density. When magnetic field is entirely
localized within the plasma, we typically find enhancement of elliptic
flow, in accordance with the  physical arguments in \cite{bv2}. However,
when the magnetic field profile extends significantly beyond the plasma
profile, plasma expansion seems to be hindered by the squeezing of
external field lines, thereby suppressing elliptic flow. Presence of
initial state fluctuations introduces extra complications due
to flux re-arrangements, as we will discuss below. It is possible that
a combination of such effects may be responsible for discrepancies between 
these various results on the expected magnetic field dependence of 
elliptic flow.  

\section{RESULTS}

We now present results of our simulations. As we mentioned, due to
small lattice size, we are able to consider evolution for a maximum of
only 3 fm time to avoid boundary effects. We first present results for
elliptic flow.

\subsection{Magnetic field dependence of elliptic flow}

 We carry out simulations with different impact parameters and
calculate elliptic flow with magnetic field and without magnetic field.
The latter is calculated by repeating the same simulation, but with 
magnetic field switched off. 

 First we present results for the conventional momentum anisotropy 
defined as $\epsilon_p = {T^{xx} - T^{yy} \over T^{xx} + T^{yy}}$. 
We calculate $\epsilon_p$ at different times with and without magnetic
field. Fig.2 shows these plots. As expected, $\epsilon_p$ increases 
gradually with time. However, we see that $\epsilon_p$ in the presence of
magnetic field increases more rapidly, clearly showing enhancement
of  momentum anisotropy due to magnetic field (for this choice of
parameters, in particular, with impact parameter of 4 fm).

\begin{figure}[!htp]
\begin{center}
\includegraphics[width=0.45\textwidth]{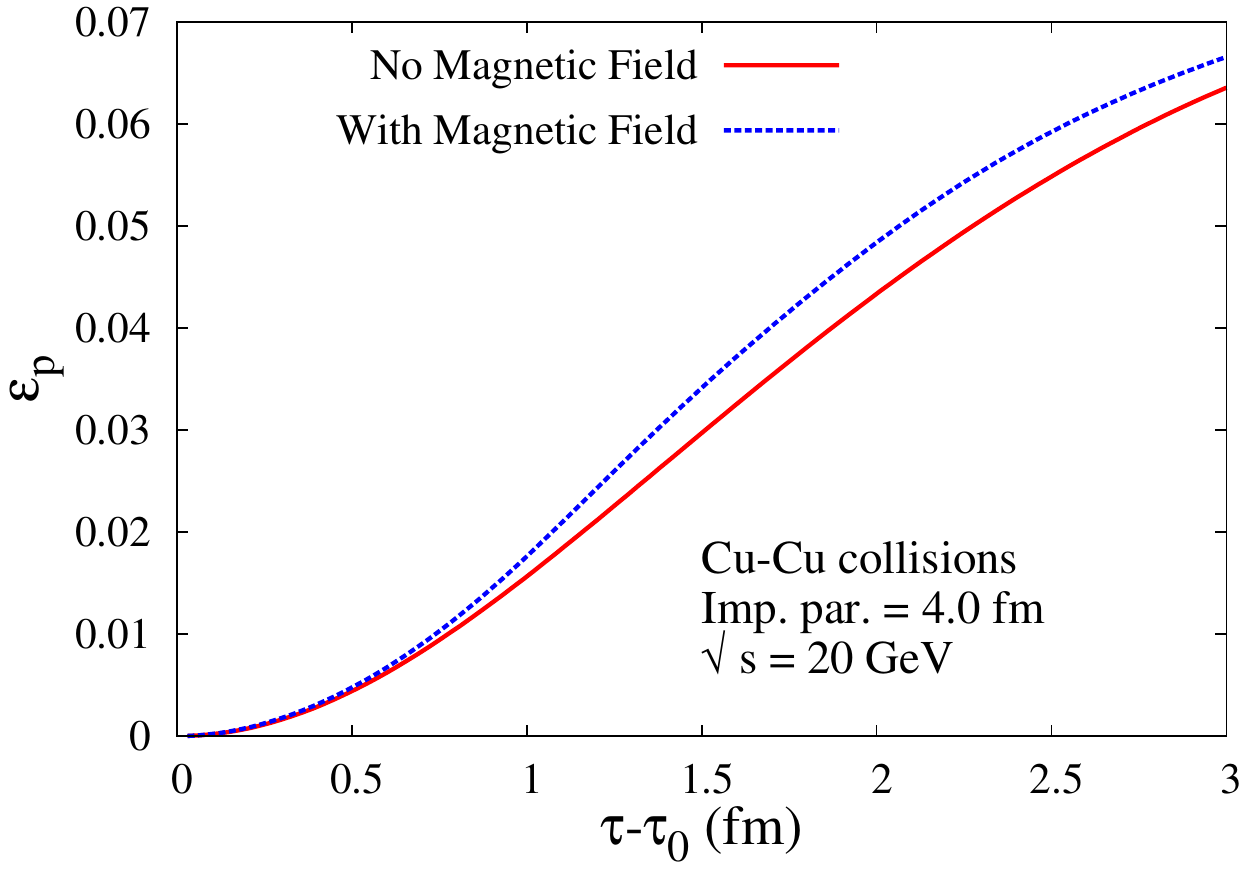}
\caption{Effect of magnetic field on build up of momentum anisotropy
$\epsilon_p$, showing clear enhancement of $\epsilon_p$ with
magnetic field, for this set of parameters, in particular for
impact parameter of 4 fm.}
\label{fig2}
\end{center}
\end{figure}

Though this expression for momentum anisotropy represents the
expected development of momentum anisotropy, we will not use this
definition of momentum anisotropy. Instead, we will use Fourier
expansion of the following normalized momentum anisotropy

\begin{equation}
f(\phi) = {\Delta p(\phi) \over {\bar p}} =
 {p(\phi) - {\bar p} \over {\bar p}}
\end{equation}

$v_2$ is taken to be the 2nd Fourier coefficient in the Fourier series
expansion of $f(\phi)$. Here $p(\phi)$ is the fluid momentum 
in a bin at azimuthal angle $\phi$ calculated from momentum components of
the energy momentum tensor, i.e. from $T^{x0}$ and $T^{y0}$, integrating
over the plasma volume in the central rapidity region of unit rapidity
width. We believe that the expression for $v_2$ obtained from Eqn.(19) is 
more appropriate as it directly gives the momentum anisotropy as measured 
in the experiment, rather than expected momentum anisotropy $\epsilon$ defined 
in terms of $T^{xx}$ and $T^{yy}$. Interestingly, this definition of
$v_2$ has a specific advantage over $\epsilon_p$. As Fig.2 shows, $\epsilon_p$
increases gradually, and becomes sizeable only after significant time
(in Fig.2 at $t = 3$ fm). So, to study effects of magnetic field on
momentum anisotropy in various conditions, it requires running simulation
every time upto significant time. In contrast, the definition in Eqn.(19)
gives a value of $v_2$ which has a large value right from the beginning
(after first few time steps), it very slowly changes afterwards due to
evolving shape of the plasma region. This may appear surprising, but
there is a simple physical explanation for this behavior. Consider a
definition of $v_2 = {T^{x0} - T^{y0} \over T^{x0} + T^{y0}}$.
(It is simple to see that the arguments given for this $v_2$ apply to
the definition of $v_2$ obtained from Fourier expansion of $f(\phi)$
in Eqn.(19).) One can see from the form of QGP energy-momentum tensor 
(Eqn.(2)) that for small velocities (at initial times), this $v_2$
equals ${v_x - v_y \over v_x + v_y}$. With initial
fluid velocity directly proportional to the pressure gradient (as
one can see from Euler's equation, see, e.g. \cite{oltr}), we
see that $v_2$ captures complete information about spatial anisotropy
right from the beginning. It does not depend on the magnitude of the
velocity, but only on the fractional difference in $v_x$ and $v_y$.
As long as the fluid acceleration remains roughly
constant, the value of $v_2$ above will remain roughly the same. 
essentially, the velocities (both $v_x$, and $v_y$, hence also fluid momenta) 
will simply increase with time. End result will be that time will not
play much role for this definition of $v_2$.
Same argument applies to $f(\phi)$ in Eqn.(19) and $v_2$ obtained from
its Fourier expansion.  That is
the reason we find that $v_2$ assumes a large, roughly constant value
right from the beginning stages, and starts changing later only with
changes in the spatial anisotropy (and effects of fluctuations etc.).
In contrast, the usual definition of momentum anisotropy $\epsilon_p$
is equal to (again, for small velocities at initial times)
${v_x^2 - v_y^2 \over v_x^2 + v_y^2 + {1 \over 2\gamma^2}}$ 
with the equation of state $\rho = 3 p_g$. This value increases from 
zero smoothly to finite value due to extra factor of ${1 \over 2\gamma^2}$ 
in the denominator as velocity magnitude  increases in time. This is why
we see $\epsilon_p$ in Fig.2 gradually increasing in time (for
both cases, with and without magnetic field).
 For our case, $v_2$ in Eqn.(19) increases rapidly to a 
finite value simply because at the first
stage itself the acceleration of the fluid (and hence the instantaneous
velocity) completely originates from the anisotropy of pressure
gradient arising from the spatial anisotropy. We find little change
in the value of $v_2$ for significant initial time (of order 2-3 fm), and
after that it evolves primarily because of the changes in the spatial
anisotropy, as expected. Thus, we believe it is much more appropriate
to use the expression for $v_2$ obtained from Eqn.(19) rather than the usual one
based on $T^{xx}$ and $T^{yy}$. This also helped us in collecting results
for many runs with different impact parameters, with and without magnetic
field, as the initial $v_2$ was itself found to be close to the time
evolved value of $v_2$ up to several fm time.

Fig.3 shows the effect of magnetic field on elliptic flow. Top figure
in Fig.3 shows the plot of $v_2(B)/v_2(0)$ vs. the impact parameter. 
We see clear enhancement in $v_2$ due to magnetic field which reaches
a peak value at the impact parameter of about 3 fm, decreasing subsequently.
Interestingly, for large impact parameter (near about 6.5 fm) there is
no effect of magnetic field on $v_2$ and for larger impact parameters,
magnetic field actually leads to suppression of $v_2$, with suppression
being strong for impact parameter of 8 fm. The bottom figure in Fig.3
shows the behavior of $v_2$ for the cases of without magnetic field
(solid, red curve) and with magnetic field (dashed and dotted curves) 
separately, clearly showing that for large impact parameters, magnetic 
field strongly suppresses the elliptic flow. 
This is despite the fact that the magnetic field is 
monotonically increasing function of the impact parameter almost for the 
entire range considered, as can be seen in Fig.4, with only slight 
decrease for the case $B_{time} = 0.4$ fm (that cannot account for
the decrease of $v_2(B)$ which is seen for both values of
$B_{time} = $ 0.4 and 0.6 fm). We will discuss below
the physical reasons for this behavior which will also explain the
discrepancies in the results of refs.\cite{bv2,tuchinv2} and ref.
\cite{mhdv2}. In all the figures, we
typically give two curves labeled by $B_{time}$ which is the time
at which the magnetic field profile is calculated after
the collision. Smaller value of $B_{time}$ gives larger value of
the magnetic field, but may not be very realistic in view of finite
conductivity of the plasma. 

\begin{figure}[!htp]
\begin{center}
\includegraphics[width=0.45\textwidth]{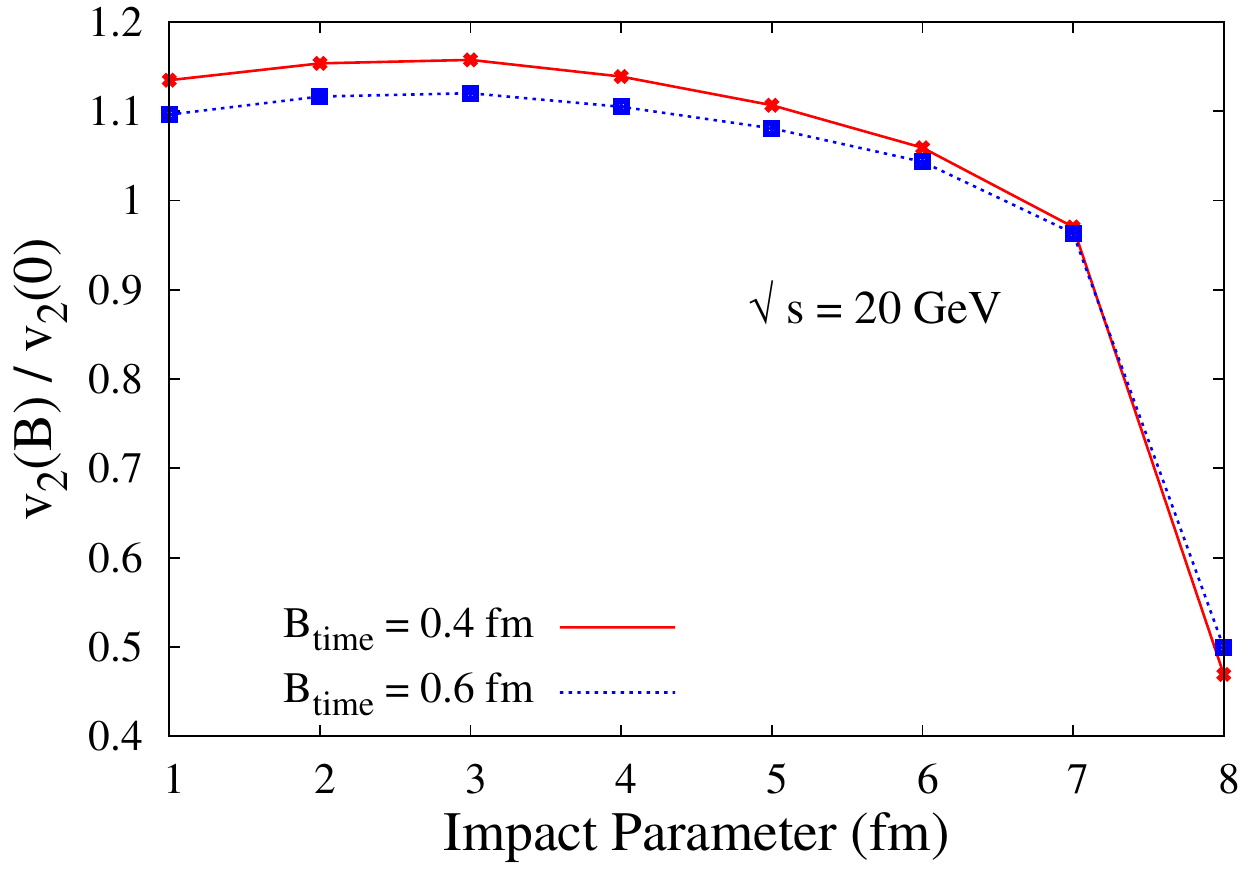}
\includegraphics[width=0.45\textwidth]{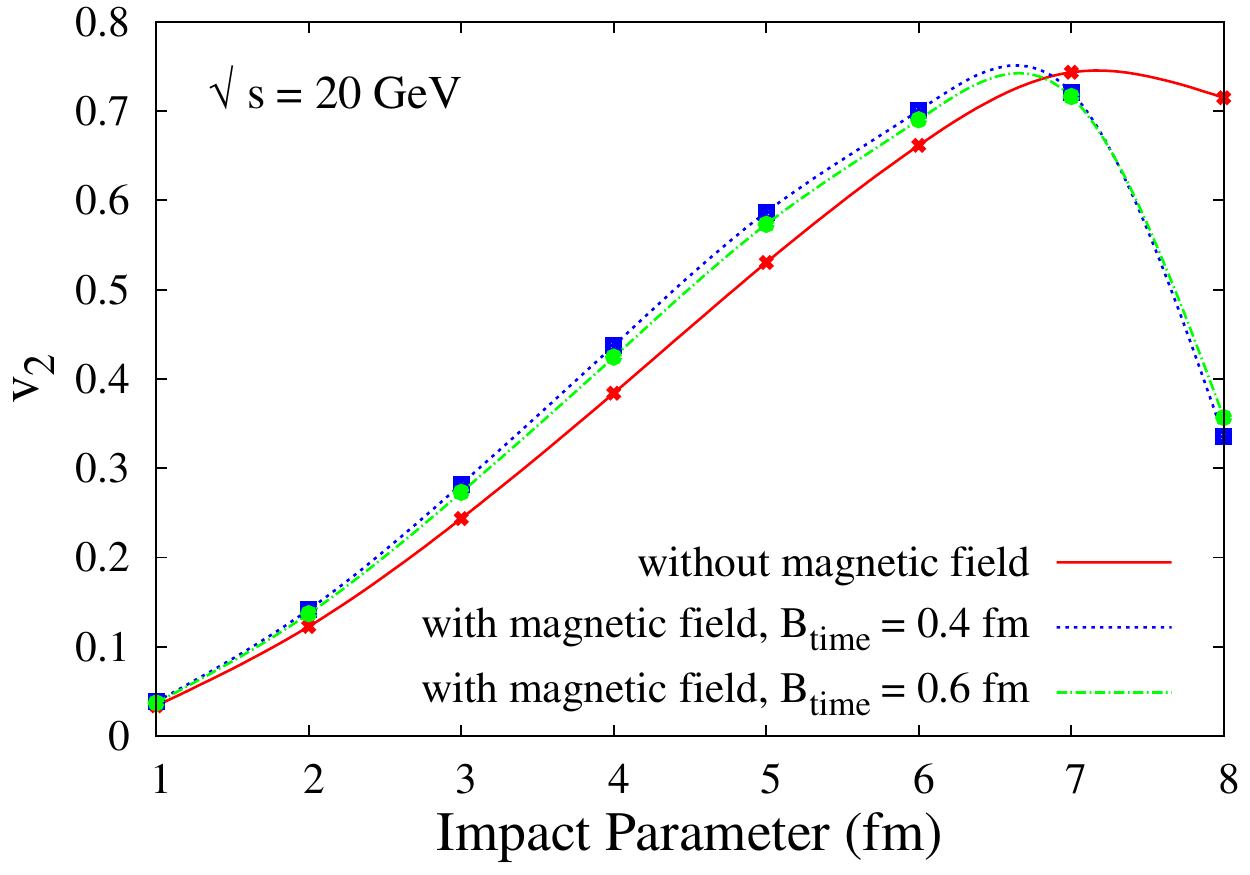}
\caption{Top figure shows the plot of $v_2(B)/v_2(0)$ vs the impact 
parameter. The ratio peaks at the impact parameter of about 3 fm, 
decreasing afterwards, and actually becomes less than 1 (meaning 
suppression of elliptic flow due to magnetic field) for large impact
parameters. Bottom figure shows the plots of $v_2$ for the cases 
of without magnetic field (solid,red, curve) and with magnetic field 
(dashed and dotted curves) separately, clearly showing that for large 
impact parameters, magnetic field strongly suppresses the elliptic flow.} 
\label{fig3}
\end{center}
\end{figure}

\begin{figure}[!htp]
\begin{center}
\includegraphics[width=0.45\textwidth]{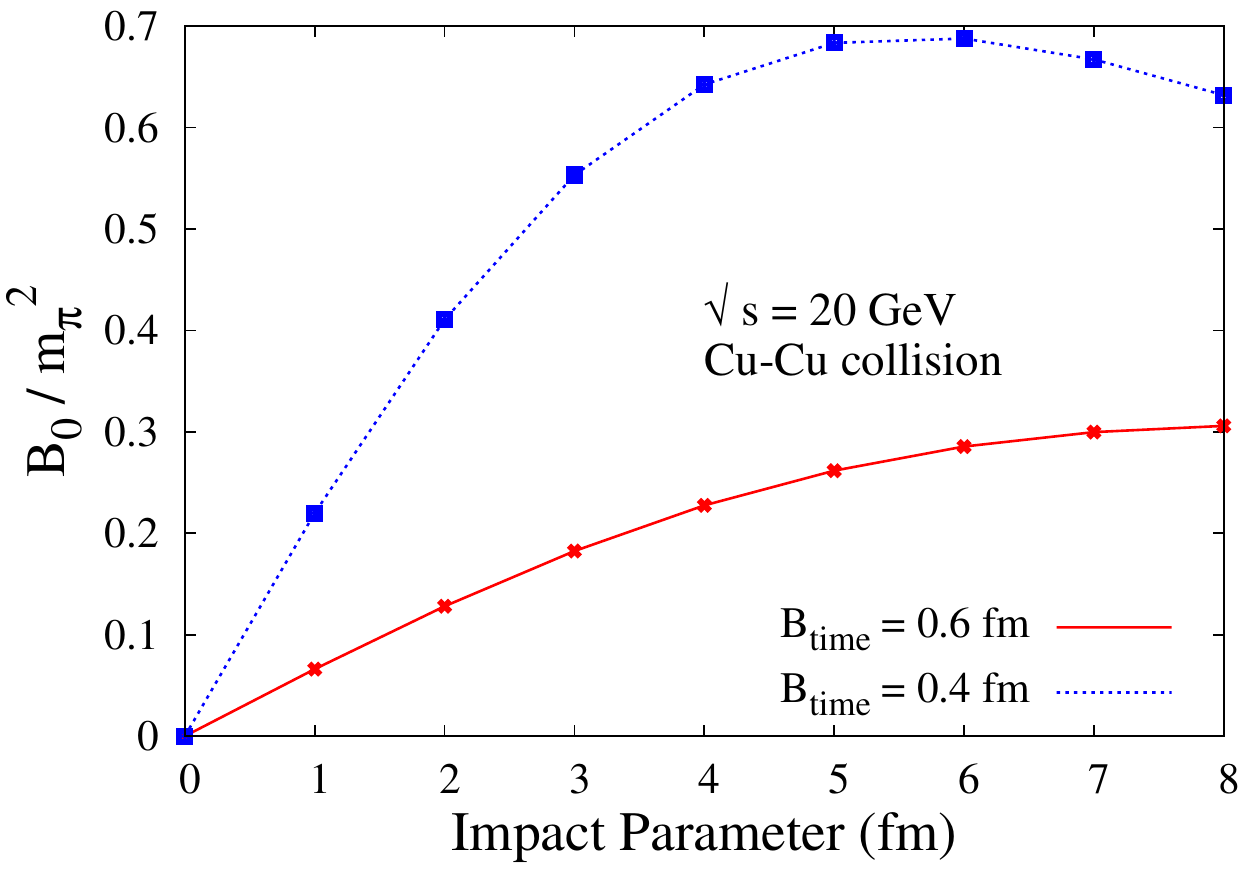}
\caption{Central value of magnetic field as a function of the impact parameter.
Note that magnetic field almost monotonically increases 
with the impact parameter.}
\label{fig4}
\end{center}
\end{figure}

 We have studied the reason for this non-trivial behavior of magnetic
field dependence of elliptic flow and it appears to originate from
the differences in the profiles of magnetic field vs. the energy density
profile. For smaller values of impact parameters, the magnetic field
profile is reasonably confined while the plasma density profile
extends for larger regions. This is the regime where arguments in
\cite{bv2,tuchinv2} seem to be valid and enhancement of $v_2$ is
seen in the presence of magnetic field. This situation is shown in 
Fig.5 which shows the initial profile  of the magnetic field as well as the
initial energy density profile for impact parameter of 1 fm.

\begin{figure}[!htp]
\begin{center}
\includegraphics[width=0.6\textwidth]{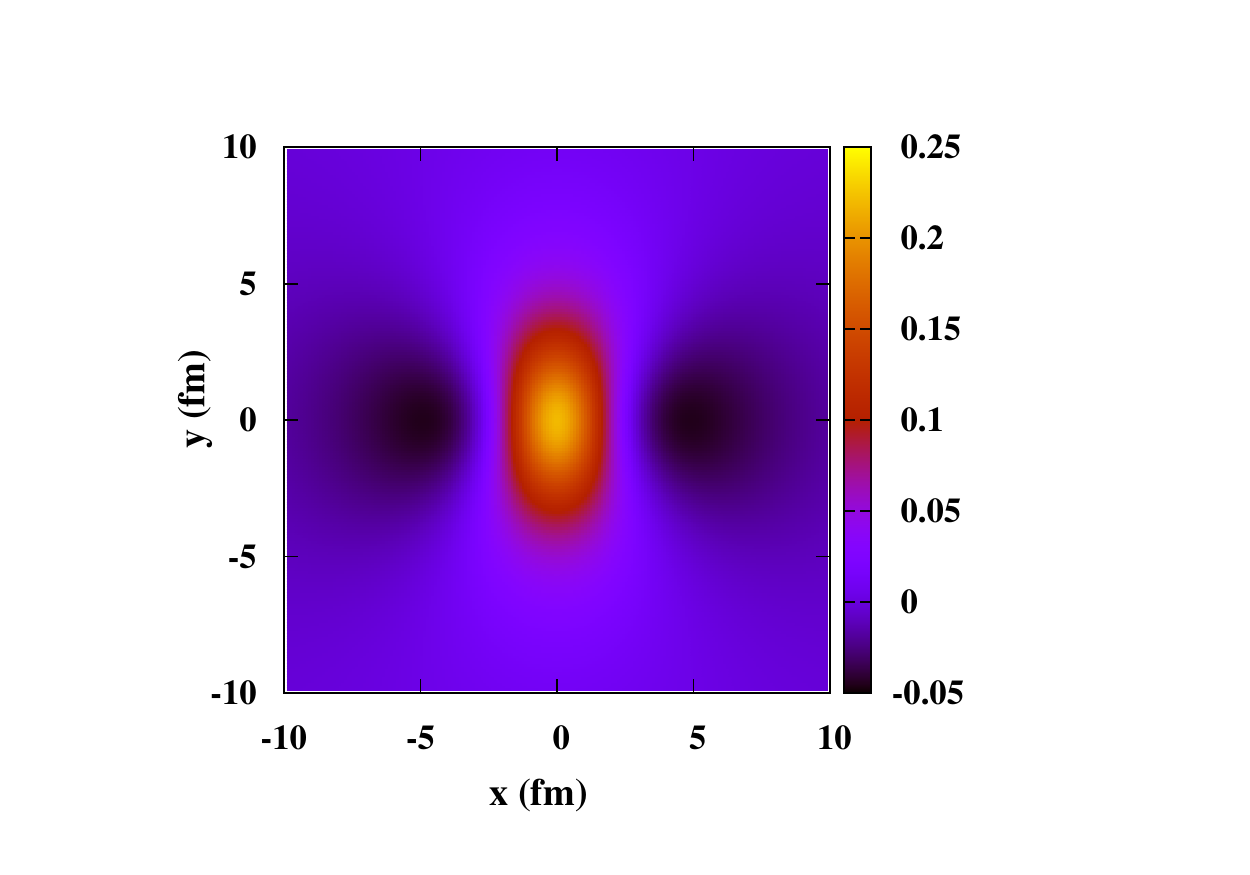}
\includegraphics[width=0.6\textwidth]{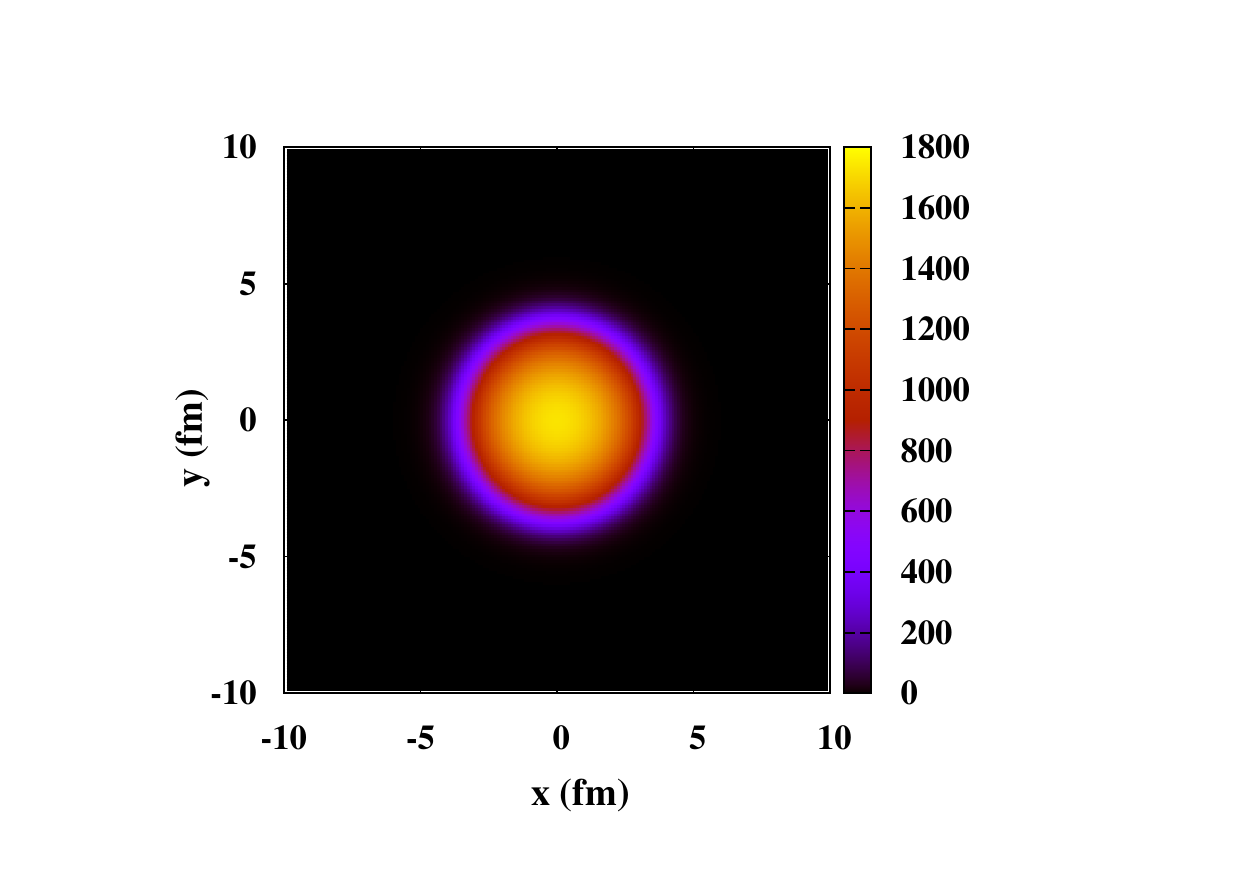}
\caption{Top figure shows the initial magnetic field profile for impact 
parameter of 1 fm. Bottom figure shows the initial plasma energy density 
profile for the same case. Note that for this small value of impact 
parameter, plasma extends well beyond the region along x-axis where 
magnetic field is significant. Here and in Fig.6 we show the y component
of the magnetic field (in the units of $m_\pi^2$, the energy density in 
the units of MeV/$fm^3$).}
\label{fig5}
\end{center}
\end{figure}

Quite opposite profiles are seen in Fig.6 which shows  initial profiles 
for magnetic field and energy density for a large impact parameter of 7
fm. (For both Figs.5,6 we have used $B_{time} = 0.4$ fm.)
Extension of significant strength of 
magnetic field profile beyond the plasma profile along x axis (semi-minor
axis of the elliptical QGP shape) squeezes plasma expansion in x-direction
as magnetic field lines in the outer regions offer stiffness against
distortion. This seems to be the cause of decrease in $v_2(B)/v_2(0)$
for larger impact parameters. This is especially demonstrated by the very 
strong decrease in $v_2(B)$ for impact parameter beyond 7 fm in the bottom 
figure in Fig.3.  (Note in this context, that simulations
in \cite{mhdv2}, where no effect of magnetic field was found on the
elliptic flow, were carried out for Au-Au collisions with large impact 
parameter.). 

\begin{figure}[!htp]
\begin{center}
\includegraphics[width=0.6\textwidth]{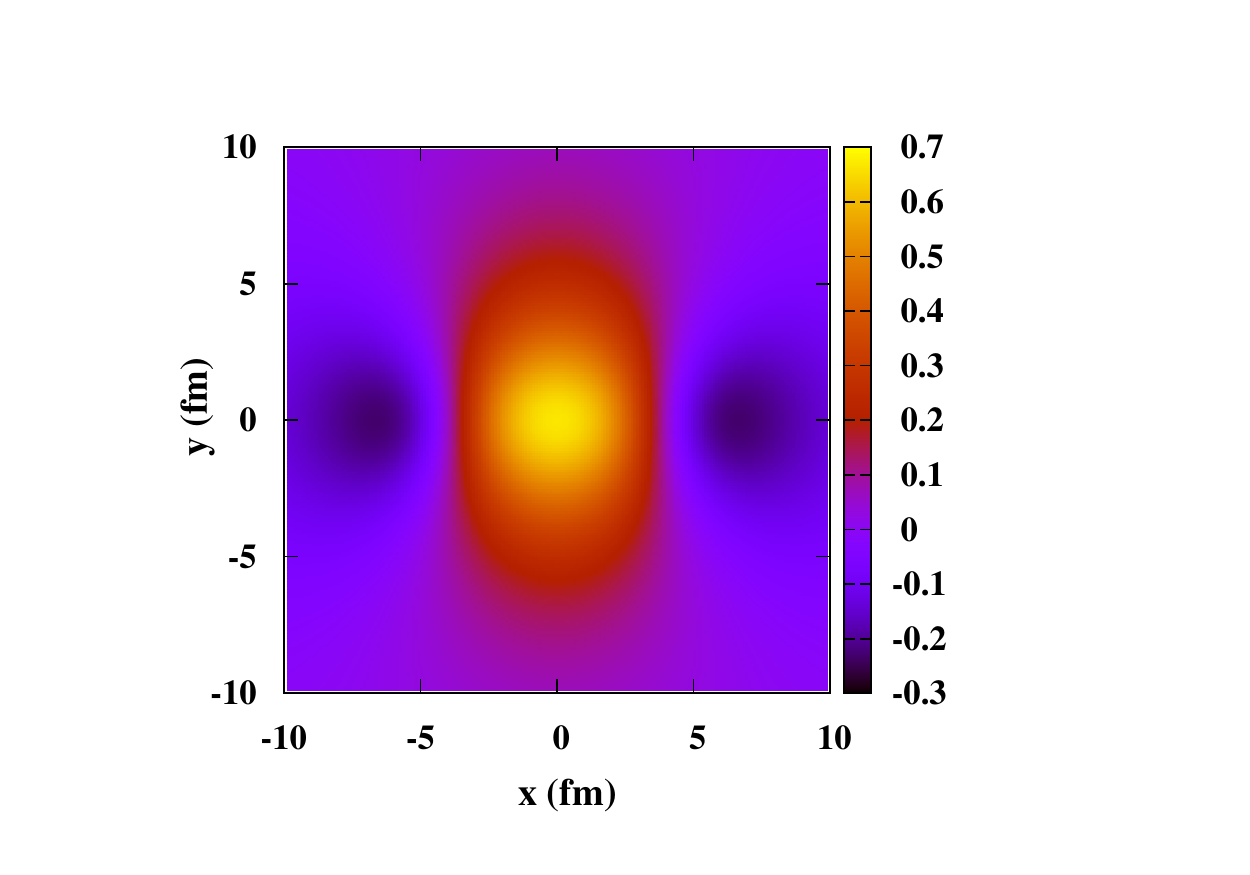}
\includegraphics[width=0.6\textwidth]{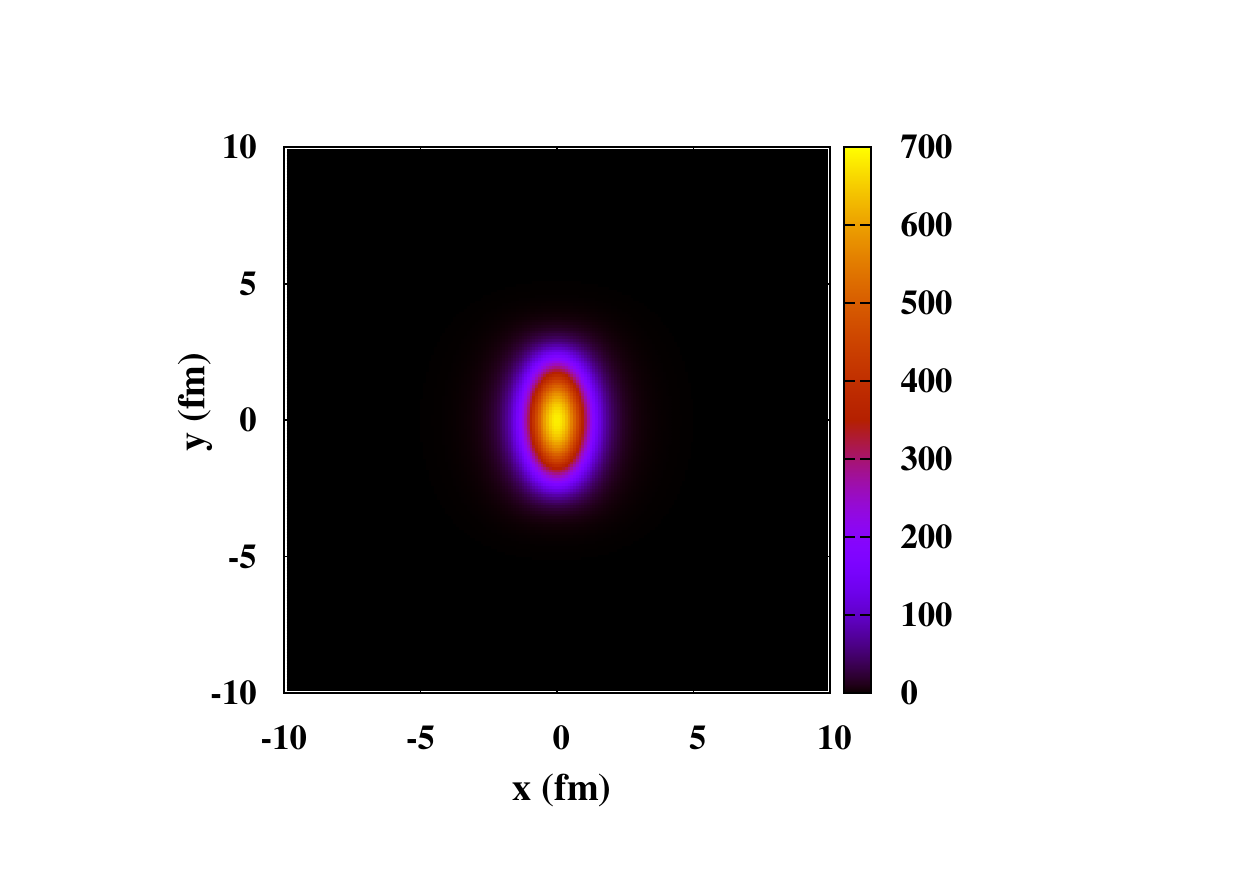}
\caption{Top figure shows the initial magnetic field profile for impact 
parameter of 7 fm. Bottom figure shows the initial plasma energy density 
profile for the same case. Note that for this large value of impact 
parameter, the two profiles show opposite behavior compared to Fig.5. 
Here we see that the magnetic field profile extends beyond the region 
along x-axis compared to the plasma energy density profile.}
\label{fig6}
\end{center}
\end{figure}

Our conclusion of this investigation is that the effect of magnetic field
on elliptic flow is quite complex. There are several physical effects
at play here, from anisotropic sound speed due to magnetic field
direction (which tends to increase elliptic flow), to Lenz's law
which suppresses plasma expansion in the regime of external magnetic
field (which tends to suppress elliptic flow). Net effect on the
elliptic flow depends on which factors dominate. We are not attempting
to provide a definitive answer to the discrepancies between different
results for $v_2(B)/v_2(0)$ in the literature, but pointing out
possible factors which may be responsible for this. Nonetheless, the
strong suppression of elliptic flow in the presence of magnetic field
for large impact parameters may provide a signal for
the presence of strong magnetic field during early stages  of
plasma evolutions. 

\subsection{Magnetic flux re-arrangement due to fluctuations}

One usually expects that magnetic field decreases as plasma evolves.
It is indeed true at an average level. However we know that the
plasma has strong initial state fluctuations in the energy density.
As fluctuations evolve, the dynamics of magnetic flux lines (which
are mostly frozen in the plasma) become very complex. It is
clearly possible that in some region plasma expansion dilutes the
magnetic flux, while due to energy density inhomogeneities, the
neighboring region may get concentration of magnetic flux, thereby
locally increasing the magnetic field. We find that indeed this happens.
Fig.7 shows the plot of central magnetic field for two different cases.
The thin curve (with stars) shows the case for Gaussian width of 0.3 fm
for the energy deposition in each binary collision in Glauber Monte Carlo, 
while the thick curve (with solid squares) represents the case of Gaussian
width of 0.4 fm, thereby representing a much smoother background
for the plasma. We see that for this smoother plasma case, the magnetic 
field roughly monotonically decreases with time (after a very little
initial increase, again due to relatively small fluctuations) as expected.
However, for the case of smaller Gaussian width, representing
stronger fluctuations, the magnetic field initially  increases significantly
almost by about 10\%, and eventually decreases.  This is only a sample
case, and it is clear that for stronger fluctuations, one may expect even
stronger temporary increase of the magnetic field during plasma evolution.
This can have important consequences for effects like chiral magnetic 
effect and chiral vortical effect (with a possibility that complex flow
pattern arising from magnetic field in the presence of fluctuations can 
in principle lead to generation of vortices).
These effects strongly depend on the presence of topological charge
density (for chiral magnetic effect) and vorticity (for chiral vortical
effect). These quantities are reasonably localized, and if the magnetic
field in these {\it relevant} regions tends to increase in time
(for some time) it can lead to strong enhancement of these effects
compared to the usual expectation based on decreasing magnetic field. 

\begin{figure}[!htp]
\begin{center}
\includegraphics[width=0.45\textwidth]{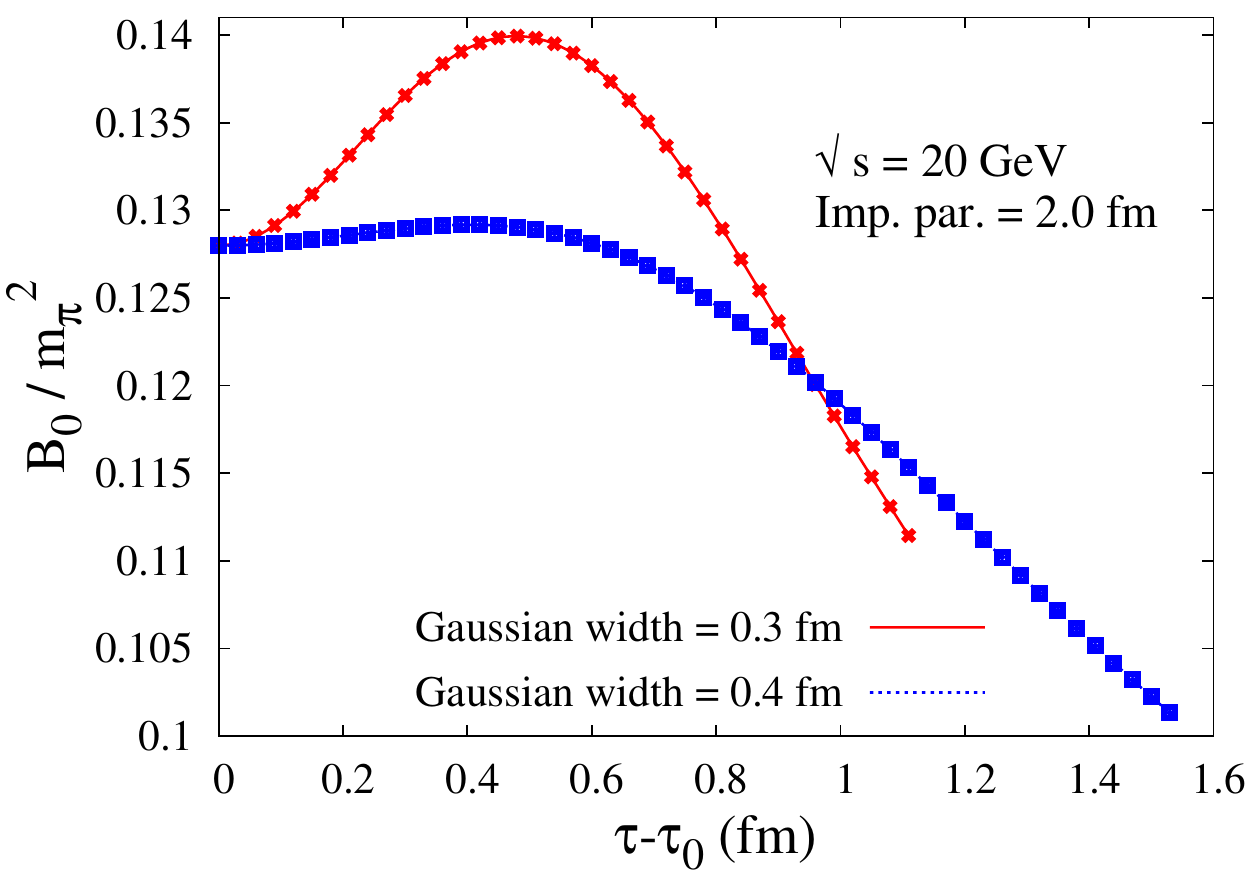}
\caption{Plot of central magnetic field in the presence of
fluctuations and for relatively smoother plasma back ground. We see 
that for the smoother case, the magnetic field monotonically decreases 
as expected. However, for the case of stronger fluctuations, the magnetic 
field initially increases, and eventually decreases.}
\label{fig7}
\end{center}
\end{figure}

\subsection{Effects of magnetic field on the power spectrum of flow
fluctuations}

 We now consider the effects of magnetic field on the power spectrum of flow
fluctuations. Power spectrum of flow fluctuations for a large number of
flow coefficients can be a very valuable
source for investigating early stages of plasma evolution 
\cite{cmbhic}. The reason for departure
from conventional focus on only first few even flow coefficients
was the recognition that initial state fluctuations contribute to
development of all flow coefficients (including the odd ones) even
for a central collisions. Many subsequent investigations confirmed this
expectation \cite {flpwr} and indeed now one routinely
measures odd coefficients (e.g. the triangular flow coefficient $v_3$) and
there have also been several investigations of power spectrum of flow
coefficients upto a large value of $n$ of about 10-12. From
the discussion above it is obvious that magnetic field will affect
the power spectrum in non-trivial manner. Indeed, it was an earlier
calculation of effects of primordial magnetic field on CMBR power
spectrum \cite{cmbrB} which prompted some of us to explore the possibility 
of magnetic field effects on the power spectrum of flow fluctuations
in RHICE \cite{bv2}. 

We use same methods for calculating flow anisotropies
as in our earlier work \cite{cmbhic}. $v_n$ denotes the
$n_{th}$ Fourier coefficient of the resulting
momentum anisotropy in ${\delta p}/p$. We do not calculate the average
values of the flow coefficients $v_n$, instead we calculate root-mean
square values of the flow coefficients $v_n^{rms}$. Further, these
calculations are performed in a lab fixed frame, without any
reference to the event planes of different events. Average values of
$v_n$ are zero due to random orientations of different events.
As $v_n^{rms}$ will have necessarily non-zero values, physically useful
information will be contained in the non-trivial shape of the power
spectrum (i.e. the plot of $v_n^{rms}$ vs. $n$).  We show below in Fig.8 
the effects of magnetic field on the power spectrum calculated after time 
evolution of about 2 fm. These results are
for realistic magnetic field for the collision energy considered here 
($\sqrt{s} = $ 20 GeV) for copper nuclei with central field strength of 0.1 
$m_\pi^2$ and 0.4 $m_\pi^2$ corresponding to $B_{time} = 0.6$ and 0.4
fm respectively (with initial state fluctuations). (For the results 
for the power spectrum calculations, we have taken  initial longitudinal 
velocity of the plasma to be zero for the stability of the program in the 
presence of strong fluctuations.) As we can see,
the effects of magnetic field are very tiny, though they are clearly
present. As we will see below, the effects of magnetic field are
not seen prominently here due to the effects of fluctuations 
being dominant for the power spectrum. Limited particle statistics may
make it very difficult to observe such tiny effects.

\begin{figure}[!htp]
\begin{center}
\includegraphics[width=0.45\textwidth]{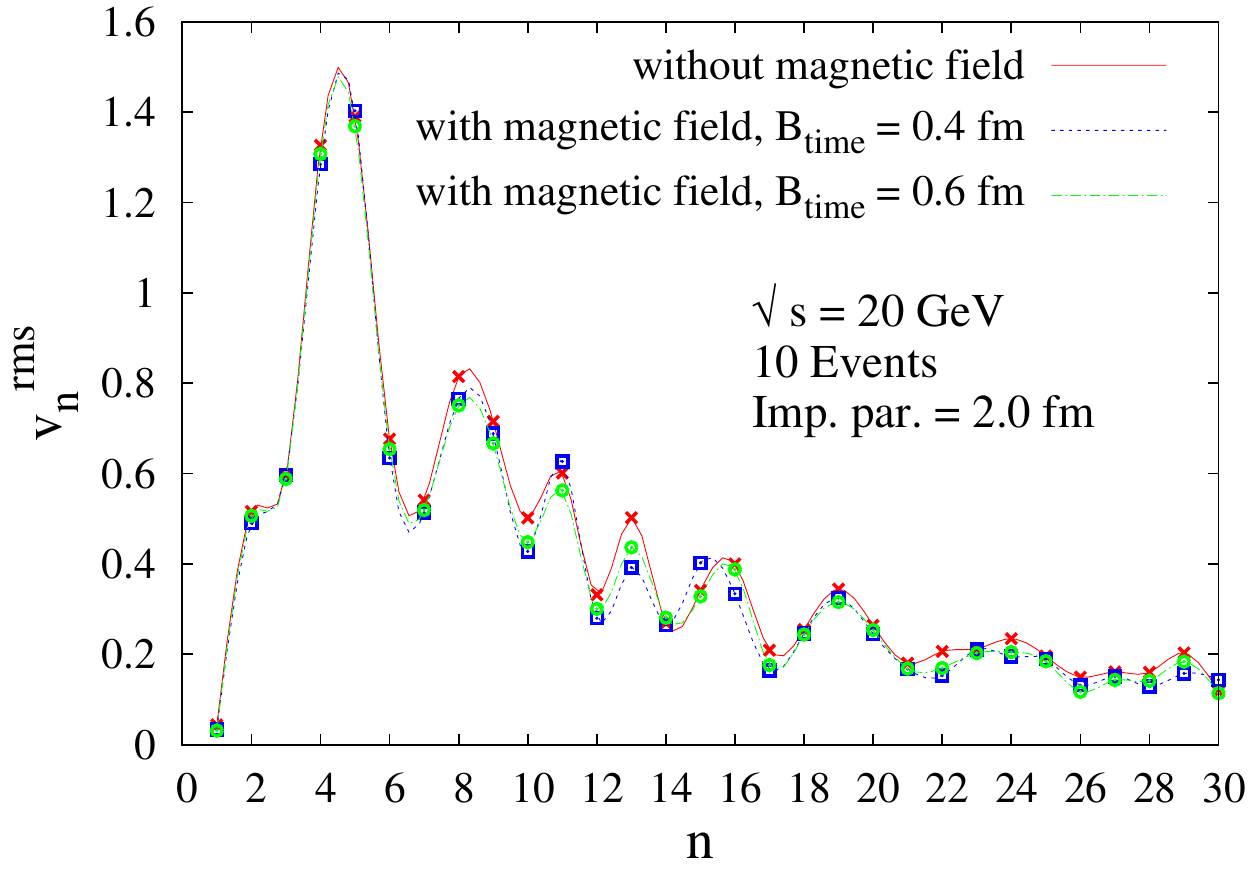}
\caption{Plot of $v_n^{rms}$ with and without magnetic field. Even though
magnetic field affects the power spectrum, its effects are very tiny here
for the magnetic field considered here (0.1 and 0.4 $m_\pi^2$).}
\label{fig8}
\end{center}
\end{figure}

 As we mentioned in the Introduction, it is of great importance to find
signals which can indicate the presence of strong magnetic field during
the initial stages. Fig.8 shows possible effects of magnetic field, 
though the effects are very insignificant for these low magnetic fields
(for much larger field appropriate for large values of $\sqrt{s}$, e.g. at
LHC, these effects may become significant. We are not able to carry out
simulations for such large values of $\sqrt{s}$ at present.) Further,
the effects seen in Fig.8 do not show any qualitatively distinct pattern
for the power spectrum. We show qualitatively different
result below for very strong magnetic fields. We consider magnetic field
strength to be $5 m_\pi^2$ and $15 m_\pi^2$. These values are completely
unrealistic here (unless unexpected things happen, say for deformed nuclei), 
and we use these only to show how completely new
effects can arise for very large magnetic field. 
As we mentioned in Sect.III, for
large magnetic fields, requiring large Lorentz gamma factor, the realistic
magnetic field profile (as used for Fig.8) causes problems with simulation.
Thus for these cases (for Figs.9,10,11 below), we use a simpler profile for 
magnetic field where the profile in the (x-z) plane is chosen to be 
proportional to the energy density profile in the (x-z) plane at y = 0.
The peak value of the magnetic field is chosen by hand. The magnetic 
field is then taken to be constant along the y axis, as consistent with 
the Gauss's law.  Fig.9 and Fig.10  below show the power 
spectrum for magnetic field of strength $5 m_\pi^2$ and $15 m_\pi^2$
respectively. As these are runs for very strong magnetic field, simulation
could  be carried out only for relatively short time of 0.6 fm.
We see strong pattern of different powers in even and odd 
$v_n^{rms}$ coefficients. This is expected from the reflection symmetry 
about the magnetic field direction if initial state fluctuations are not 
dominant.  Note that for $5 m_\pi^2$ case, even-odd pattern
is seen for only first few flow coefficients as fluctuation effects wash out
the effect for larger $v_n$ for the event average over 10 events. 
For $15 m_\pi^2$ case the magnetic field is very strong and fluctuation 
effects are not able to wash out the even-odd pattern arising from the
magnetic field. This is a qualitatively distinct
result and can give unambiguous signal for the presence of
strong magnetic field during early stages.

\begin{figure}[!htp]
\begin{center}
\includegraphics[width=0.45\textwidth]{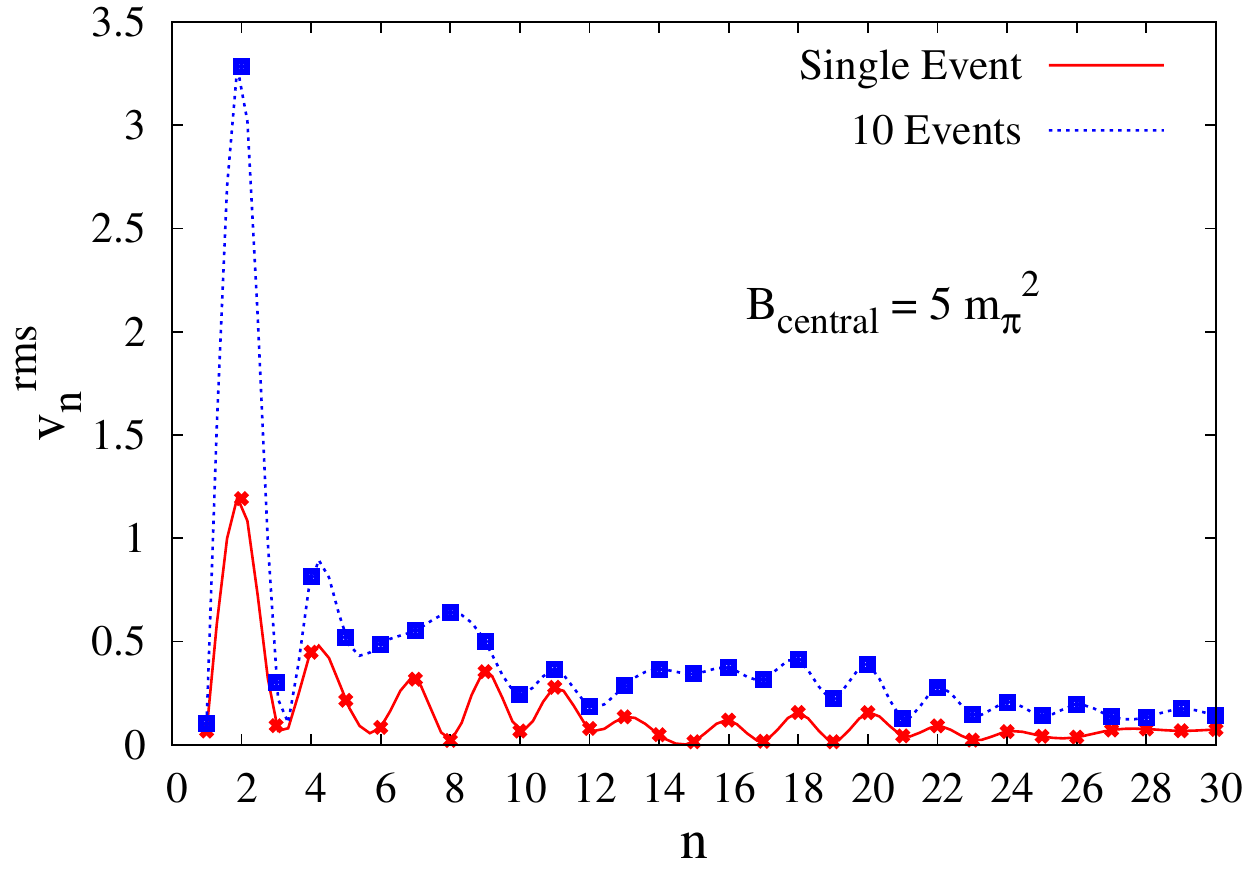}
\caption{Plot of $v_n^{rms}$ for magnetic field with
strength $5 m_\pi^2$. Even-odd power difference is seen in first
few flow coefficients as fluctuations wash out the effect for
large $v_n$s.}
\label{fig9}
\end{center}
\end{figure}

\begin{figure}[!htp]
\begin{center}
\includegraphics[width=0.45\textwidth]{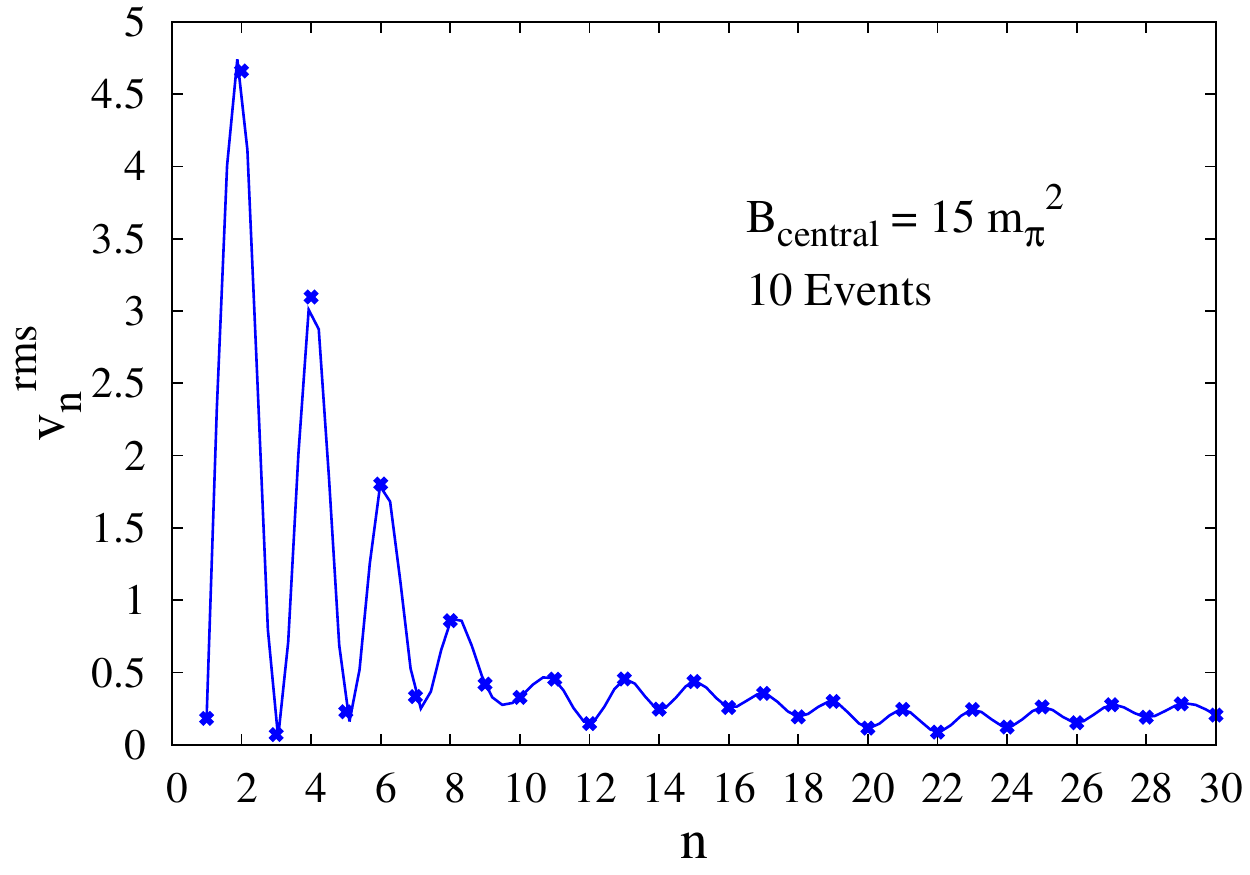}
\caption{Plot of $v_n^{rms}$ for very strong magnetic field with
strength $15 m_\pi^2$. Strong difference in the power of even and odd
values of $v_n^{rms}$ are seen. Though such large magnetic field are completely
unexpected here, such effects  may provide unambiguous signal 
for the presence of any unexpected strong initial magnetic field.}
\label{fig10}
\end{center}
\end{figure}

 The reason one needs very strong magnetic field is that although
magnetic field tends to develop clear pattern of even-odd power
difference, there are strong effects of initial state fluctuations 
on the power spectrum. The final power spectrum is a combined effect
of the two patterns. Strong magnetic field is needed to dominate
over the effects of fluctuations in Fig.9,10. To illustrate this,
we show in Fig.11 flow fluctuations for a smooth isotropic plasma
region (without any fluctuations) in the presence of magnetic field.
We now take a more reasonable value of magnetic field strength
equal to $m_\pi^2$. Due to smaller magnetic field and smooth
plasma profile, the evolution could be run up to 3 fm time (after which
boundary effects could not be neglected). We see that strong even-odd 
power difference is present in the power spectrum.

\begin{figure}[!htp]
\begin{center}
\includegraphics[width=0.45\textwidth]{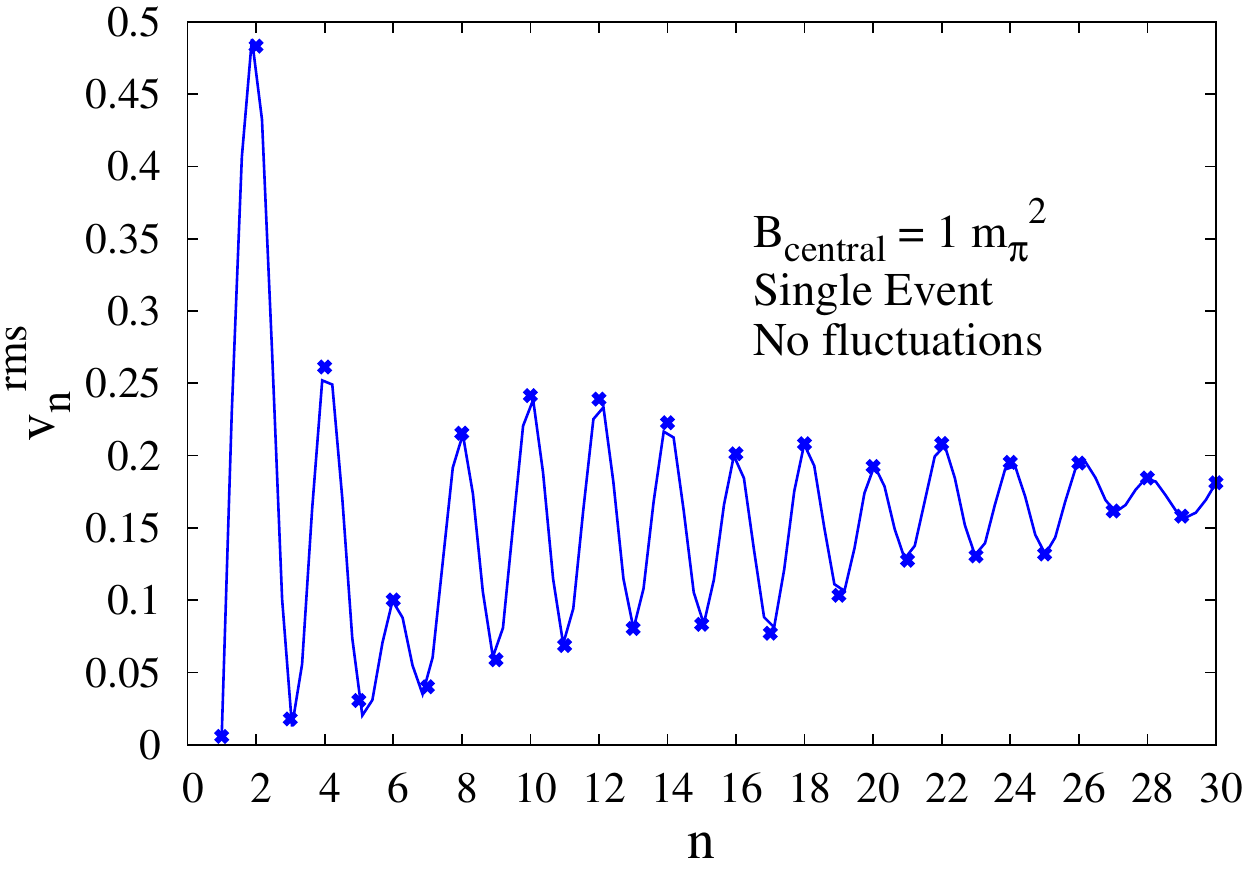}
\caption{Plot of $v_n^{rms}$ for magnetic field with
strength $m_\pi^2$. Here we consider isotropic region with smooth
plasma profile without any fluctuations.  Strong difference in 
the power of even and odd values of $v_n^{rms}$ are present
arising from the effect of magnetic field.}
\label{fig11}
\end{center}
\end{figure}

We mention that such even-odd power difference can arise due to
presence of vortices also during early plasma evolution, as 
demonstrated in our earlier work \cite{qcdsf}. Thus, we may conclude
that even-odd difference in the power spectrum indicates either
strong magnetic field or presence of vortices in the initial plasma.
(We know that to some extent the effect of magnetic field in a plasma
is similar to the presence of vortices as the Lorentz force due to
magnetic field has similar form as the Coriolis force in the presence of 
vortices.)  This result also has interesting implications for the CMBR power 
spectrum. It is known that low $l$ modes of CMBR power spectrum also show 
possible difference in even-odd modes \cite{evenodd}. It is possible 
that this feature may be indicative of the presence of a magnetic field, 
or presence of some vorticity, during the very early stages of the inflation.

\subsection{Anomalous elliptic flow for deformed nucleus}

 Collision of deformed nuclei  opens up entirely new range of
possibilities for heavy-ion collisions. This is especially true
when considering possible magnetic field configurations for
a given shape of plasma. For non-central collisions of 
spherical nuclei, one is constrained to consider the magnetic field 
pointing along the semi-major axis of the elliptical QGP region.
(Though due to fluctuations, deviations from this will happen
but roughly the picture remains the same.) For deformed nuclei,
entirely new possibilities can arise. As an example, Fig.12 shows 
ellipsoidal nuclei, with longer axes of both along the y axis, with
impact parameter also along the y axis. As one can see from Fig.12,
different impact parameters can lead to following anomalous magnetic
field configurations (in the sense that they cannot arise for spherical
nuclei).

a) QGP region being elliptical in shape but the magnetic field
pointing along the semi-minor axis, x-axis in this case as seen in 
Fig.12a.

b) QGP region being roughly spherical, but still strong magnetic field
is present due to strong components coming from spectators, as seen in 
Fig.12b.

\begin{figure}[!htp]
\begin{center}
\includegraphics[width=0.5\textwidth]{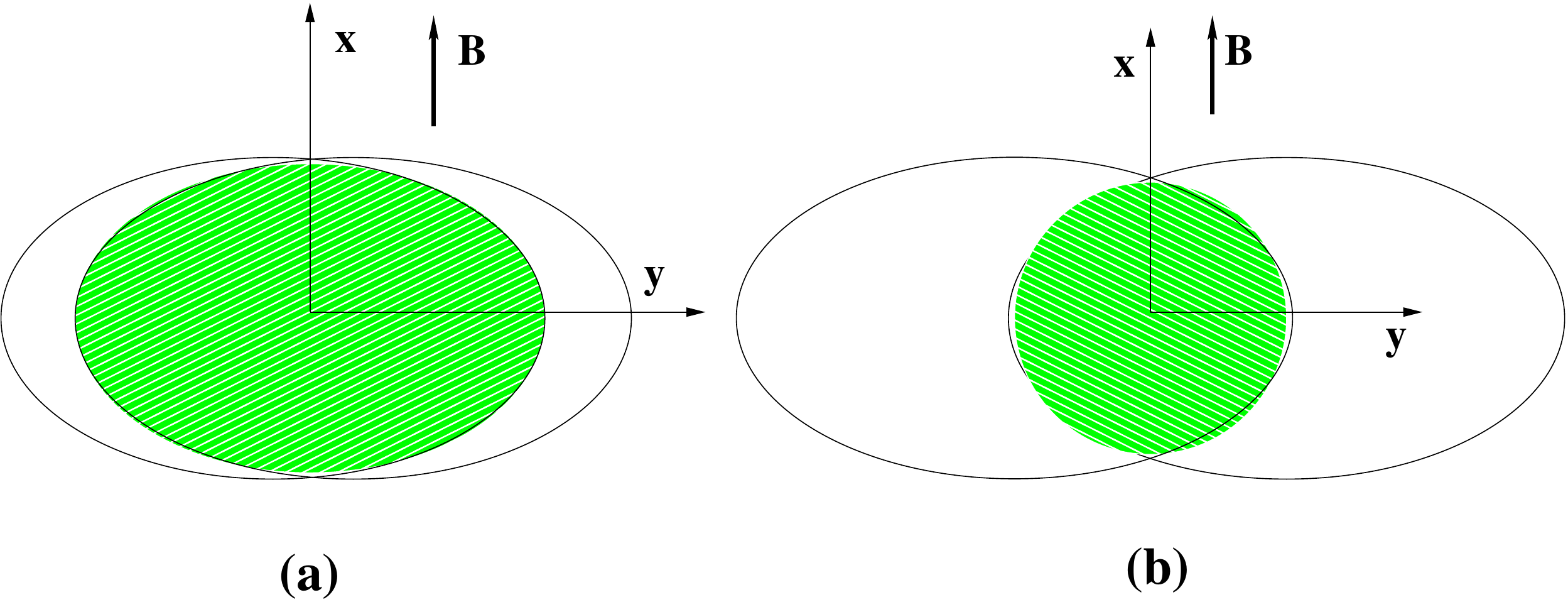}
\caption{(a) shows the situation of the case when the
QGP region is elliptical in shape but the magnetic field
points along the semi-minor axis. (b) shows the case 
when the QGP region is roughly spherical, but still strong magnetic field
is present due to strong components coming from spectators.}
\label{fig12}
\end{center}
\end{figure}

With the physics of effects of magnetic field as described above,
one can immediately guess what to expect in both these cases. For
(a) we expect suppression of elliptic flow as the magnetic field induced
anisotropy leads to larger momentum flowing in the direction of 
semi-major axis of the elliptical QGP shape, even though the usual fluid
pressure gradient develops larger flow along the semi-minor axis.
This leads to strong suppression of elliptic flow due to this anomalous
magnetic field. (For very strong magnetic field the net $v_2$ may
even  be completely dominated by the magnetic field, leading to negative 
elliptic flow.) For (b) one would have expected no elliptic
flow for the smooth plasma profile considered here, (non-zero $v_2$
may only arise from any fluctuations), as the QGP is
roughly isotropic. However, the presence of strong magnetic field
introduces anisotropic pressure, leading to development of
non-zero $v_2$, even though QGP region is spherical. Figs.13,14 
confirm these expectations. 
Again, the anomalous elliptic flow in these situations may provide a 
signal for initial stage magnetic field.

Note that here we are not simulating collision
of deformed nuclei. We use the plasma profile for Fig.12a and Fig.12b
by using collisions of spherical nuclei (copper) with non-zero
and zero impact parameter respectively. But for the magnetic field
we calculate the magnetic field as in Sects.5A and 5B, rotate it along
the x axis, and use that for
the evolution of the above plasma profiles. This, in some sense, models
different situations of collisions of deformed nuclei as in Fig.12a,b.
A full simulation for deformed nuclei is presently under investigation
and will be presented in a future work. 

\begin{figure}[!htp]
\begin{center}
\includegraphics[width=0.45\textwidth]{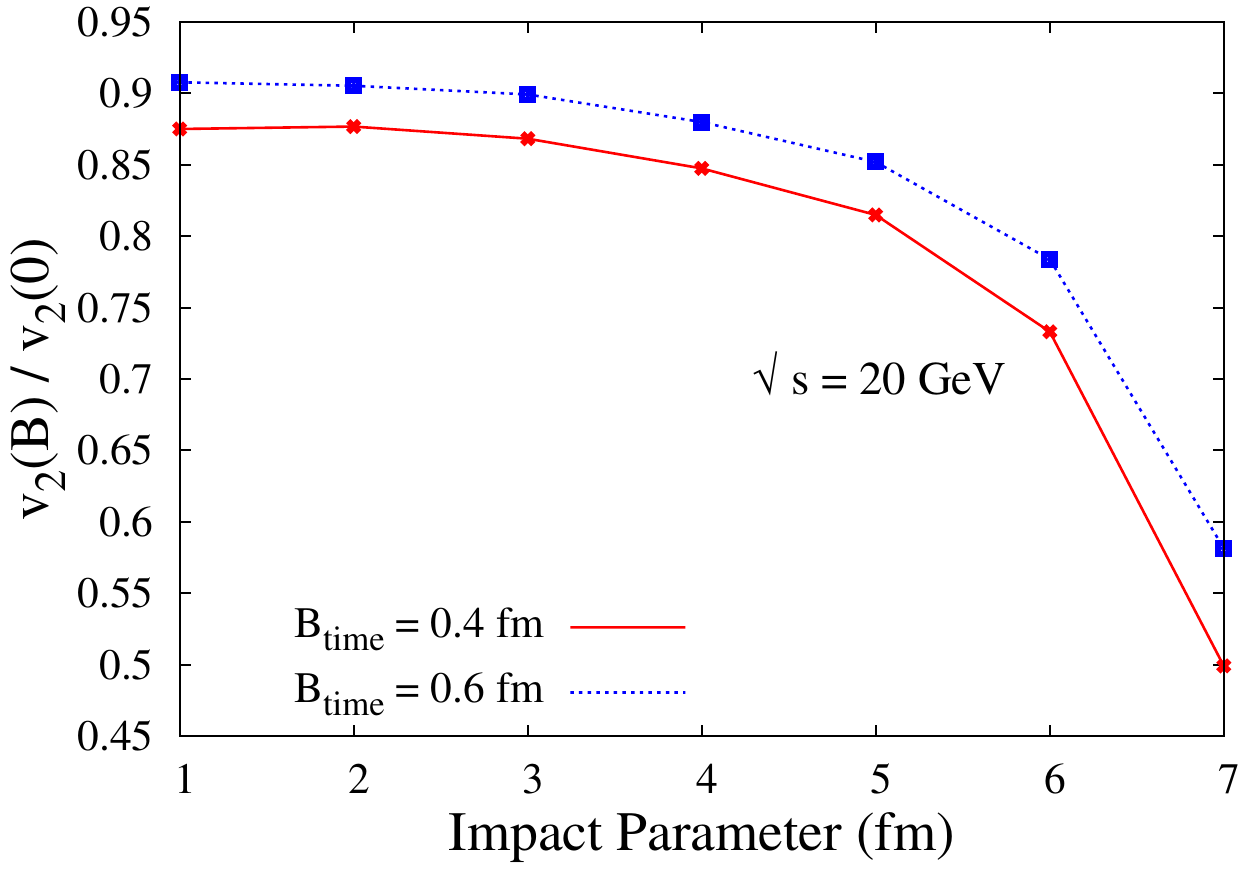}
\caption{$v_2$ for the case when the QGP region is elliptical in shape 
but the magnetic field points along the semi-minor axis, x-axis in 
this case. Strong suppression of elliptic flow arises with larger
momentum flowing in the direction of semi-major axis of the elliptical 
QGP shape due to magnetic field induced anisotropy.}
\label{fig13}
\end{center}
\end{figure}

\begin{figure}[!htp]
\begin{center}
\includegraphics[width=0.45\textwidth]{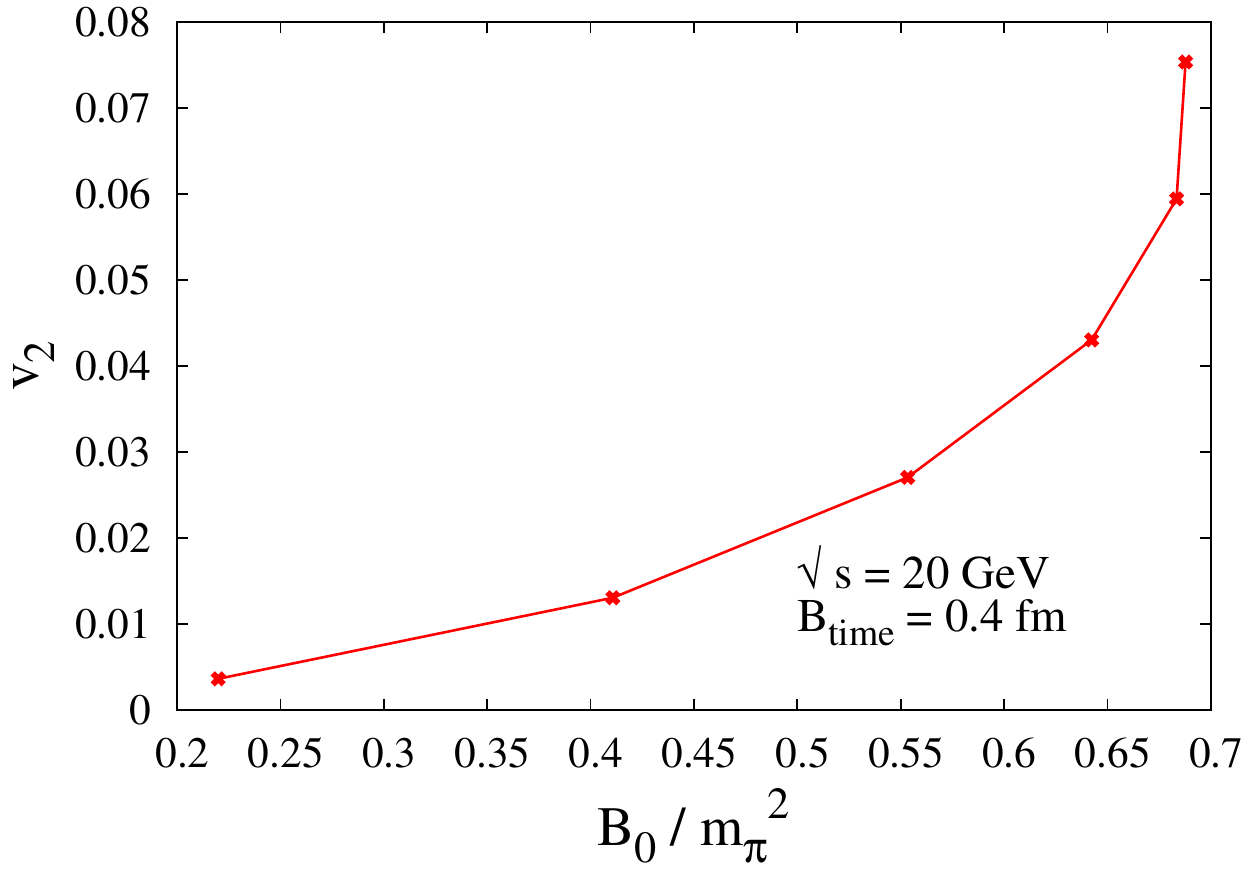}
\caption{$v_2$ for the case when the QGP region is roughly isotropic
in shape but still non-zero magnetic field is present leading to
non-zero $v_2$, monotonically increasing with the strength of
magnetic field, even though no elliptic flow is expected
in this case.}
\label{fig14}
\end{center}
\end{figure}

\subsection{Quadrupole magnetic field from deformed nucleus}
 
 A very interesting possibility arises when considering collision of
deformed nuclei. Consider again ellipsoidal nuclei with long
axes in the transverse plane (as in the above), but now in crossed
configuration. Fig.15 shows this crossed configuration for Uranium nucleus
with the semi-minor and semi-major axes being  about 6.7 fm and 8.7 fm
respectively. Magnetic field is calculated at time of 0.4 fm
after the collision for $\sqrt{s} = 20$ GeV. It is clear
that while the resulting QGP region is roughly isotropic (possibly
with strong $v_4$ component), spectators will generate quadrupolar
magnetic field as one can see from Fig.16 showing the magnetic field
lines for this crossed configuration of colliding nuclei. 
(Magnetic field here has been calculated by extending the calculation
of Sec.V A,B for the case of deformed nucleus, Uranium in this case.
We calculate magnetic field from uniformly charged ellipsoidal nuclei
\cite{dfrm,jcksn}, oppositely moving, with appropriate Lorentz transformations.)
This raises very important possibilities. Quadrupolar field
will itself contribute to $v_4$, thereby affecting final value of $v_4$
of the plasma.  Further, quadrupolar field will tend to focus plasma motion 
along the longitudinal direction, thereby affecting Bjorken longitudinal
expansion itself. For charged plasma with finite conductivity one
may expect charge separation in the transverse direction as a function of 
rapidity, while a focusing effect should be seen along the beam axis for
the plasma. This should lead to suppression of
transverse flow at non-zero rapidity. Further, if focusing is strong,
it may lead to hot extended regions along the longitudinal axis.  
This requires a detailed investigation of plasma dynamics with such
a crossed configuration collision of deformed nuclei properly represented
in Glauber Monte Carlo. This  is under study and will be presented
in a future work.

\begin{figure}[!htp]
\begin{center}
\includegraphics[width=0.22\textwidth]{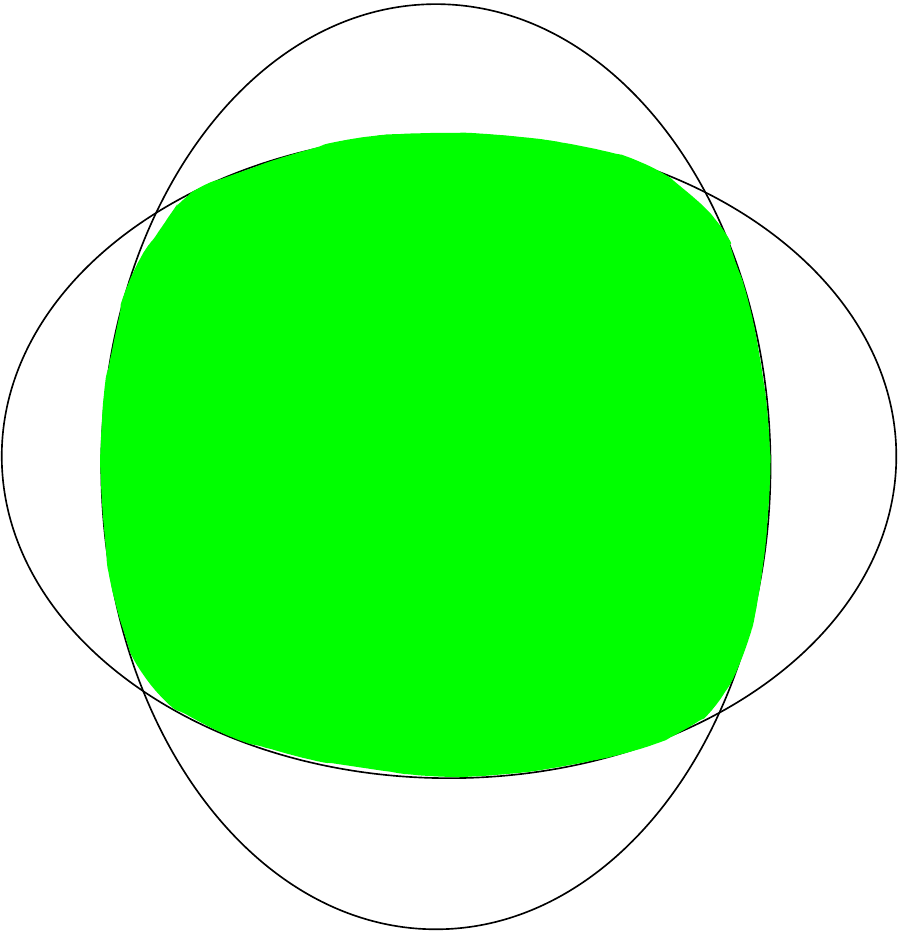}
\caption{Crossed configuration of collision of deformed nuclei.
Note that the overlap region will be reasonably isotropic,
with possibly strong $v_4$ component. Importantly now there are
four spectator parts whose motion should lead to quadrupolar
magnetic field configuration.}
\label{fig15}
\end{center}
\end{figure}

\begin{figure}[!htp]
\begin{center}
\includegraphics[width=0.45\textwidth]{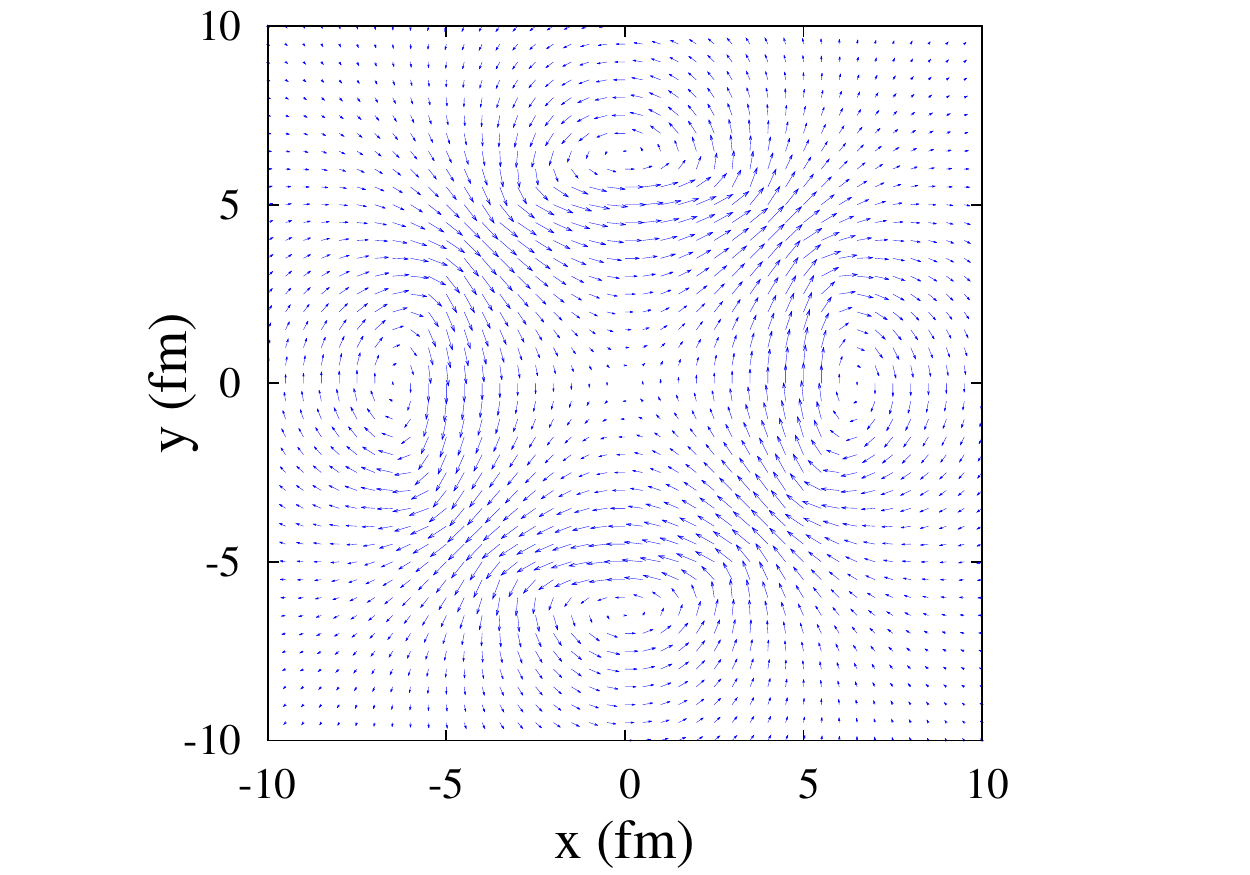}
\caption{Magnetic field configuration arising from collision of
crossed deformed nuclei (Uranium) as in Fig.15. Quadrupolar nature 
of the field is clear.}
\label{fig16}
\end{center}
\end{figure}

\section{Conclusions}

We have demonstrated qualitatively new effects on the flow pattern of
QGP in the presence of initial magnetic field. As we emphasized, due to
various limitations of our simulation,
we are not in a position to provide numbers which can be compared to
experimental data. Our intention is to show possibilities of new physical
phenomena which one can try to look for in experiments. These qualitative
patterns may be able to  provide clear signal for the presence of strong 
magnetic field during early stages of the evolution, though actual value
of magnetic field etc. will depend on more reliable numerical estimates 
of the numbers. Among our results one of the results
shows that due to flux re-arrangement arising from evolving fluctuations,
there may be local regions where magnetic field increases for some time
(before it starts decreasing finally). If topological charges or
vortices are also present in that region, it can lead to enhancement
of chiral magnetic/vortical effects. We see very complex patterns of 
twisting flow developing due to magnetic field effects in the presence of
fluctuations. For strong fluctuations and strong magnetic field, it
seems entirely possible that localized vorticity may get generated at
later times which we are not able to study due to limitations of
our simulation.  Our result on enhancement of elliptic
flow in the presence of magnetic field confirms earlier expectation
in refs.\cite{bv2,tuchinv2}. At the same time our simulation also
points out that the effects of magnetic field on 
elliptic flow are much more complex than envisaged in simple arguments
of ref.\cite{bv2}. In fact in some situations one finds decrease
in the elliptic flow. This may resolve the discrepancy between the results
of ref.\cite{bv2,tuchinv2} and ref.\cite{mhdv2} 
(see, also refs.\cite{bv2ref1,bv2ref2}). The strong suppression of elliptic
flow for large impact parameters can provide a signal for initial
stage strong magnetic field. (For this it is needed to have observations
extended for very large impact parameters, to distinguish from the
suppression from usual hydrodynamics resulting from decreased plasma
pressure at large impact parameters.) We show non-trivial
effects of magnetic field on the power spectrum of flow fluctuations.
The strongest form of this effect being in the form of
even-odd power difference in the flow power spectrum for strong magnetic 
fields which can be a very clean signal for strong magnetic field, or 
vortices \cite{qcdsf}, in RHICE. (At the same time, it can have important 
implications for the low $l$ modes for CMBR power spectrum.) Our results 
for deformed nuclei provide possibilities of anomalous elliptic flow, 
which can be used to detect the magnetic field in such collisions.
It points to a very interesting possibility of generating a quadrupolar 
magnetic field configuration which can have focusing effect on plasma
in the longitudinal direction (along with a possibility of charge
separation in the transverse direction.) These possibilities are under
investigation at present and will be presented in a future work.
 
\acknowledgments

  We are very grateful to Partha Bagchi, Srikumar Sengupta, Nirupam Dutta,
Oindrila Ganguly, Ranjita Mohapatra, and Ananta P. Mishra for useful 
comments and discussions. Some of the results here were presented by AMS 
in the conferences  {\it QCD Chirality Workshop 2015} held at Physics Dept. 
UCLA (USA), in Jan. 2015,  and in March 2017, and at the conference 
{\it Hadronic Matter under Extreme Conditions} held at JINR (Russia) 
in Nov. 2016, by AD in the conference {\it Quark Matter 2017} at Chicago, 
Feb. 2017, and by SSD at the conference {\it XQCD-2017} at Pisa, Italy,
June 2017. We thank the participants in these meetings for very useful
comments and suggestions.



\begin{thebibliography}{99}

\bibitem{cp} D. Kharzeev, R.D. Pisarski, and M.H.G. Tytgat,
{\it Phys. Rev. Lett.} {\bf 81}, 512 (1998);
S.A. Voloshin, {\it Phys. Rev. C} {\bf 70}, 057901 (2004).

\bibitem{cp2} D.E. Kharzeev, L.D. McLerran, and H.J. Warringa,
{\it Nucl. Phys. A} {\bf 803} 227 (2008).

\bibitem{bv2}  R. K. Mohapatra, P. S. Saumia, and A. M. Srivastava,
Mod. Phys. Lett. {\bf A 26}, 2477 (2011).

\bibitem{etabys} P. Romatschke and U. Romatschke, Phys. Rev.
Lett. {\bf 99}, 172301 (2007).

\bibitem{tuchinv2} K. Tuchin, J.Phys. G39, 025010 (2012). 

\bibitem{mhdv2} G. Inghirami et al. arXiv: 1609.03042; Eur. Phys.
J. {\bf C 76}, 659 (2016).

\bibitem{bcal1} M. Asakawa, A. Majumder, and B. Muller,
{\it Phys. Rev. C} {\bf 81}, 064912 (2010).

\bibitem{tuchin} K. Tuchin, Phys.Rev. {\bf C 83}, 017901 (2011); 
{\it Phys. Rev. C} {\bf 82}, 034904 (2010).

\bibitem{dfusn} L. Landau and E. Lifshitz, {\it Electrodynamics of
Continuous Media} (Pergamon Press, N.Y., USA, 1984), Sect. 58.

\bibitem{jcksn} J.D. Jackson, {\it Classical Electrodynamics}, 
3rd Edition (John Wiley \& Sons, Inc., USA, 1999).

\bibitem{sigma} H.-T. Ding et al.  arXiv:1012.4963, Phys. Rev. {\bf D 83},
034504 (2011).

\bibitem{bref} K. Tuchin, Phys. Rev. {\bf C 88}, 024911 (2013).

\bibitem{sigmaB} P.V. Buividovich, et al., {\it Phys. Rev. Lett.} {\bf 105},
132001 (2010).

\bibitem{mhd} A. Mignone and G. Bodo, Mon. Not. R. Astron.
Soc. {\bf 368}, 1040 (2006).

\bibitem{gbmc} M.L. Miller, K. Reygers, S.J. Sanders, 
and P. Steinberg, Ann. Rev. Nucl. Part. Sci. {\bf 57}, 205 (2007);
P.F. Kolb, J. Sollfrank, and U. W. Heinz, Phys.Rev. {\bf C 62}, 054909 (2000).


\bibitem{mhdwave} J.P. (Hans) Goedbloed, R. Keppens, and S. Poedts,
{\it Advanced Magnetohydrodynamics} (Cambridge University Press, UK, 2010).

\bibitem{oltr} J.-Y. Ollitrault, {\it Eur. J. Phys.} {\bf 29}, 275 (2008);
R. S. Bhalerao, J. P. Blaizot, N. Borghini, and
J.-Y. Ollitrault, {\it Phys. Lett. B} {\bf 627}, 49 (2005).

\bibitem{bv2ref1} L.G. Pang, G. Endrodi, and H. Petersen, Phys.
Rev. {\bf C 93}, 044919 (2016).

\bibitem{bv2ref2} S. Pu and D.L. Yang, Phys. Rev. {\bf D 93}, 054042 (2016).

\bibitem{cmbhic}  A. P. Mishra, R. K. Mohapatra, P. S. Saumia, and
A. M. Srivastava, Phys. Rev. {\bf C 77}, 064902 (2008);
Phys. Rev. {\bf C 81}, 034903 (2010).

\bibitem{flpwr} P. Sorensen, J.Phys. G 37 (2010) 094011;
B. Alver and G. Roland, Phys. Rev. {C 81} (2010) 054905;
A. Mocsy and P. Sorensen, Nucl. Phys. {\bf A 855} (2011) 241;
J.I. Kapusta, Nucl. Phys. {A 862-863} (2011) 47;
J.Y. Ollitrault, J. Phys. Conf. Ser. 312 (2011) 012002;
P. Sorensen, arXiv:0808.0503.

\bibitem{cmbrB} J. Adams, U.H. Danielsson, D. Grasso, and
H. Rubinstein, {\it Phys. Lett. B} {\bf 388}, 253 (1996).

\bibitem{qcdsf} A. Das, S.S. Dave, S. De, and A.M. Srivastava,
arXiv: 1607.00480, to appear in Mod. Phys. Lett. A.

\bibitem{evenodd} P.K. Aluri and P. Jain, Mon. Not. Roy. Soc. 
{\bf 419}, 3378 (2012).

\bibitem{dfrm} O.D. Kellogg, {\it Foundations of Potential Theory},
(Springer-Verlag, 1967).

\end{thebibliography}
\end{document}